\documentclass[lettersize,journal]{IEEEtran}
\usepackage{amsmath,amsfonts}
\usepackage{algorithmic}
\usepackage{algorithm}
\usepackage{array}
\usepackage[caption=false,font=footnotesize,labelfont=rm,textfont=rm]{subfig}
\usepackage{textcomp}
\usepackage{stfloats}
\usepackage{url}
\usepackage{threeparttable}
\usepackage{verbatim}
\usepackage{graphicx}
\usepackage{cite}
\usepackage{enumitem}
\usepackage{xcolor}
\usepackage{hyperref}
\usepackage{bm}

\DeclareMathOperator*{\minimize}{minimize}
\hypersetup{
colorlinks=true,
linkcolor=black,
filecolor=black,      
urlcolor=black,
citecolor=black
}

\begin{document}

\title{DiffCharge: Generating EV Charging Scenarios\\ via a Denoising Diffusion Model}


\author{Siyang Li, Hui Xiong, \emph{Fellow, IEEE}, and Yize Chen, \emph{Member, IEEE}
\thanks{This work was partially supported by National Natural Science Foundation of
China   (Grant   No.92370204),   Guangzhou-HKUST(GZ)   Joint   Funding   Program
(Grant No.2023A03J0008), Education Bureau of Guangzhou Municipality, and
Guangdong Science and Technology Department. 

The authors are with the Artificial Intelligence Thrust, Hong Kong University of Science and Technology (Guangzhou), email: sli572@connect.hkust-gz.edu.cn, xionghui@hkust-gz.edu.cn, yizechen@ust.hk.}\vspace{-15pt}
}

\markboth{Journal of \LaTeX\ Class Files,~Vol.~14, No.~8, August~2021}%
{Shell \MakeLowercase{\textit{et al.}}: A Sample Article Using IEEEtran.cls for IEEE Journals}


\maketitle

\begin{abstract}
Recent proliferation of electric vehicle (EV) charging load has imposed vital stress on power grid. \textcolor{black}{The stochasticity and volatility of EV charging behaviors render it challenging to manipulate the uncertain charging demand for grid operations and charging management.} \textcolor{black}{Charging scenario generation can serve for future EV integration by modeling charging load uncertainties and simulating various realistic charging sessions.}
To this end, we propose a denoising \underline{D}iffusion-based \underline{C}harging scenario generation model coined DiffCharge, which is capable of yielding both battery-level and station-level EV charging time-series data with distinct temporal properties. In principle, the devised model can progressively convert the simply known Gaussian noise to genuine charging demand profiles by learning a parameterized reversal of the forward diffusion process. Besides, we leverage the multi-head self-attention mechanism and prior conditions to capture the unique temporal correlations associated with battery or charging station types in actual charging dynamics. Moreover, we validate the superior generative capacity of DiffCharge on a real-world dataset involving ample charging session records, and attest the efficacy of produced charging scenarios on a practical EV operation problem in the day-ahead electricity market.
\end{abstract}

\section{Introduction}
\textcolor{black}{\IEEEPARstart{I}{n} order to reduce the carbon emissions and achieve a clean energy system, electric vehicles (EVs) have been extensively adopted. Such prominent EV penetration incurs non-negligible charging load for the distribution grid, posing huge challenges on the grid planning and operation~\cite{Lamedica2019}. The statistics show that the Western Interconnection (WECC) grid in the United States would foresee a 9-26\% rise in the peak total load with full EV adoption \cite{powell2022charging} in 2035.}
In this regard, grid operators are required to properly manage the increasing EV integration using coordinated charging strategies to both fulfill grid operation goals and meet user charging needs \cite{8521585}. However, due to the stochastic and volatile charging behaviors as well as heterogeneous battery operating characteristics like charging control protocols and maximum power limit \cite{Li2023}, it is quite hard to model the uncertainties of EV charging demand. \textcolor{black}{With the growth of fast charging facilities and strong interplay between power grids and battery charging activities, it is essential to effectively represent EV charging sessions by distinct features such as charging rate variations, start/end time and duration.}



There exist various uncertain factors in real EV charging behaviors, including battery charging protocols, arrival/departure time, battery type, state-of-charge (SOC) level and daily number of station-level charging transactions \cite{Lee2019}. The fundamental question remains how to efficiently characterize EV charging loads, so as to better inform grid operators to cope with such ever-changing and fast-increasing charging demand. 

Generating a variety of high-resolution EV charging scenarios is a promising solution. By producing a set of energy time-series data to represent all possible future trajectories, scenario generation has been employed to tackle the ubiquitous uncertainties in renewable power and residential demand \cite{Jinghua2020}. But utilizing such generative models to capture EV charging uncertainties has not been well-explored. \textcolor{black}{In both \cite{POWELL2022123592} and \cite{Quirós2018}, Gaussian Mixture Models (GMMs) are leveraged to explicitly estimate the joint distribution of daily charging load profiles. However, due to the Gaussian prior postulation and limited Gaussian components in distributional functions, the representation capacity of GMMs is restricted, and it is hard for them to recover accurate temporal dynamics in varieties of EV charging time-series data.}

In particular, EV charging time-series can be prescribed from two perspectives: \emph{battery-level} and \emph{station-level}. The battery-level scenarios comprise fine-grained charging curves of single EVs \cite{Chenxi2020}, which can be sensed by individual charging piles. Charging curves can not only be affected by the dynamical and electrochemical properties of battery states, but also moulded by a series of socioeconomic factors like commuting distance and available charging duration. \textcolor{black}{Station-level scenarios encompass diverse daily charging load profiles \cite{huang2023metaprobformer}, which are essentially the aggregation of total charging sessions served on the same day at the same site. They manifest disparate temporal patterns of overall charging demand and distinct usage modes of charging facilities. Besides, such scenarios are also affected by station-level charging limits and customer charging habits.} System operators and charging utilities can benefit from such charging temporal scenarios by designing more robust distribution grids and more reliable operating strategies to host potential large-scale EV penetration \cite{jaworski2023vehicle}.

In this work, we aim to develop a potent scenario generation model to learn such heterogeneous charging load time-series by approximating their real distribution accurately. The key challenge lies in \emph{how to capture distinctive temporal dynamics and express various temporal patterns in real charging demand profiles}. We address this issue by devising a potent denoising diffusion probabilistic model (DDPM) \cite{Ho2020} to derive both battery-level and station-level EV charging time-series scenarios. DDPM has recently demonstrated exceptional capacity to learn the high-dimensional and complex distribution of multi-modal data, as well as generating a wide range of authentic samples of high quality and mode diversity. The inspiration behind DDPM is physics-informed,
i.e. intuitively, if the perturbed noise in the diffusion process can be identified exactly, we can eliminate them (denoise) reversely to restore the initial yet unknown real data distribution, such as that for charging demand time-series. Specifically, in the forward process, a series of Gaussian noises with increasing magnitudes are gradually added to the original data distribution until it is eventually transformed into a simple normal distribution with known mean and variance. Then in the reverse process, a set of parameterized Gaussian transitions are learned to predict the imposed step-wise noise during the forward process. 

Although diffusion models have exhibited excellent ability to synthesize images and texts of high fidelity and diverse modes \cite{Ling2023}, few of works showcase their capability on time-series generation \cite{coletta2023constrained}, \cite{tashiro2021csdi}, especially to yield high-resolution EV charging temporal data. Accordingly, we extend DDPM to generate both battery-level and station-level charging time-series scenarios in this paper. 
The proposed model can produce plenty of realistic charging load time-series while well covering their various temporal patterns. \textcolor{black}{The realism of generated samples can be inspected from the view of statistical quality, i.e. the distributional similarity to historical real data \cite{coletta2023constrained} and can be assessed by a group of quantitative metrics defined in Section \ref{sec:setup}-C.} Besides, we utilize the multi-head self-attention mechanism to learn the complex temporal correlations and variations elicited by charging dynamics. The main contributions of this article are summarized as follows:
\textcolor{black}{
\begin{enumerate}[leftmargin=*]
 \item{\emph{Versatile EV charging scenario generation tasks}. We prescribe the scenario generation task from two aspects to handle EV charging uncertainties, including individual battery-level charging curves and aggregated station-level charging load profiles. Learning such two-fold time-series scenarios can capture both temporal dynamics induced by unique battery charging behaviors and diverse daily power demands from different charging stations.
}
 \item{\emph{Diffusion-based scenario generation}. We propose a data-driven denoising diffusion model termed DiffCharge, which can derive the intractable charging uncertainties and generate various charging load profiles with realistic and distinctive temporal properties. To the best of our knowledge, this is one of the pioneering work to realize the effectiveness of diffusion models on energy time-series generation.
 }
 \item{\emph{Efficient learning}. We exhibit that DiffCharge can procure the complex distribution of miscellaneous charging load profiles by simply training a denoising network to predict the step-wise Gaussian noise. And it can be easily scaled to generate different categories of EV charging time-series with merely moderate adjustments.
 }
\end{enumerate}
}

\textcolor{black}{
The proposed DiffCharge and simulation results are open-sourced at \href{https://github.com/LSY-Cython/DiffCharge}{\textcolor{black}{https://github.com/LSY-Cython/DiffCharge}}. The rest of the paper is organized as follows: Section \ref{sec:related works} describes related works and potential applications of synthetic scenarios. Section \ref{sec:formulation} specifies our goals to generate two-level charging time-series. Section \ref{sec:pre} and \ref{sec:model} clarify the principle and detailed designs of the devised framework. Section \ref{sec:setup} and \ref{sec:results} describe the experimental settings and evaluate the generation outcomes using different metrics and an pragmatic operation problem. Section \ref{sec:conclusion} concludes the paper.
}

\section{Related Works}
\label{sec:related works}
Scenario generation and probabilistic forecasts are actively utilized to handle stochasticity and intermittency in renewable energy system, which are beneficial for uncertainty-aware grid dispatch and operation \cite{BUZNA2021116337}. \textcolor{black}{Classical methods such as Markov Chain Monte Carlo (MCMC) \cite{papaefthymiou2008mcmc}, auto-regression-based \cite{Morales2010} and copula-based models \cite{Jinghua2020} rely on statistical assumptions and presumed structures to fit the sequential dependence. Such methods are hard to capture the joint distribution and complex nonlinear temporal correlations in original time-series.}

Due to the exceptional capacity to derive the complex data distribution and capture hidden temporal relations, the machine learning-based methods become prevalent, which can generate numerous realistic and diverse renewable scenarios. For instance, the generative adversarial networks (GANs) can employ adversarial training to produce renewable production profiles for multiple power plants \cite{Chen2018, DONG2022118387, yuan2022conditional}. The normalizing flows (NFs) and variational auto-encoders (VAEs) involved in \cite{DUMAS2022117871} can yield accurate photovoltaic scenarios in conformity to weather forecasts, attributed to their exact density estimation. Diffusion models are initially used in \cite{Esteban2023} to achieve time-series scenario forecasts for renewables, but the temporal and distributional characteristics are yet to be discussed. 

Although such data-driven generative models have gained excellent performance on renewable scenario generation, they have yet been proved to be competitive on generative modeling for EV charging time-series. Authors in \cite{jahangir2021novel} propose to characterize charging sessions as 3D images and employ GANs to produce a chunk of associated charging behaviors. But they only consider modeling the arrival/departure time and total requested energy for multiple EVs, and ignore the rich temporal information in real charging curves. There also exist several works seeking to model the station-level charging load profiles via statistical methods. \textcolor{black}{In \cite{POWELL2022123592, Quirós2018}, GMMs are utilized to estimate the joint distribution of multiple random variables related to charging sessions. Then, a batch of EV charging events are sampled and aggregated into daily load profiles via a two-stage battery model. Yet due to the expressivity bottleneck of GMMs, while EV charging curves are highly nonlinear and stochastic, it is difficult for GMMs to fully retain and generate complex and realistic temporal patterns in real charging data.} \textcolor{black}{The MCMC sampling method is applied in \cite{ng2015markov, papaefthymiou2008mcmc, singh2021prosumer} to identify the temporal state transitions. Nonetheless, because of the prior assumption of GMMs and rough estimation for discrete transition probabilities in MCMC, they are hard to discover the actual distribution and temporal modes hidden in real charging power time-series.}

\textcolor{black}{In this paper, we consider to cultivate a dedicated diffusion model to generate specific charging temporal scenarios. Note that diffusion models have attained excellence on generating controllable driving scenarios for autonomous vehicles (e.g. multi-agent motion trajectories), coupled with a flurry of complementary techniques containing conditioning mechanism, score-based guidance and constraint sampling \cite{jiang2023motiondiffuser, xu2023diffscene}. A much closer work \cite{dong2023short} proposes to execute day-ahead wind power profile generation via a latent diffusion model conditioned on weather covariate forecasts. \textcolor{black}{Different from these existing diffusion-based scenario generation tasks \cite{jiang2023motiondiffuser, xu2023diffscene,dong2023short}, we aim to capture stochastic and variable EV charging behaviors by yielding both individual battery-level and aggregated station-level temporal charging scenarios, rather than uncertainties in vehicle moving trajectories or renewable production. And we focus on fostering an efficient diffusion model which can derive unique temporal dynamics in two-stage battery charging rate curves as well as distinct temporal patterns in daily station charging load profiles. Whereas other existing diffusion-based generation methods are not required to preserve such detailed domain-specific temporal properties described in Section~\ref{sec:formulation}.} Moreover, diffusion-based generation can bypass the issues of training instability and mode collapse in GANs, and the barrier of bijective function design in NFs, since it evades to train auxiliary networks and only requires a denoising network to predict the imposed step-wise noise.}

\textcolor{black}{There are many potential applications for synthetic time-series scenarios by DiffCharge. On the one hand, by modeling real-world EV charging curves, we can obtain the complicated distribution of battery charging dynamics and exploit it to simulate the underlying battery electrochemical processes, largely reducing the hurdle to identify equivalent circuit models \cite{Chenxi2020}. We can also learn to monitor battery health conditions or predict future charging rates on top of such supplementary realistic curves \cite{tian2021deep}. On the other hand, both charging service and distribution grid operators can benefit from such charging uncertainty modeling. For example, generative models trained on historical charging data can be applied to simulate a wide variety of unseen scenarios which can elicit more efficient stochastic optimization for EV charging scheduling \cite{Cui2019}, electricity market bidding \cite{jahangir2020plug} and route planning \cite{Lin2022}. Besides, different modes of likely charging load profiles are crucial for operators to analyze the impact of charging stations on grid status \cite{Leou2014}, and for planners to determine the proper charging infrastructure capacity along with coordinated distributed generation and energy storage \cite{Cui2019}. Moreover, the meta-learning method can be used to transfer the probabilistic forecasting ability forged on other real charging load distributions to newly built charging stations with few session records \cite{huang2023metaprobformer}.}

\section{Problem Formulation}
\label{sec:formulation}

\begin{figure}[t]
\centering
\setlength{\abovecaptionskip}{0.cm}
\includegraphics[width=0.48\textwidth]{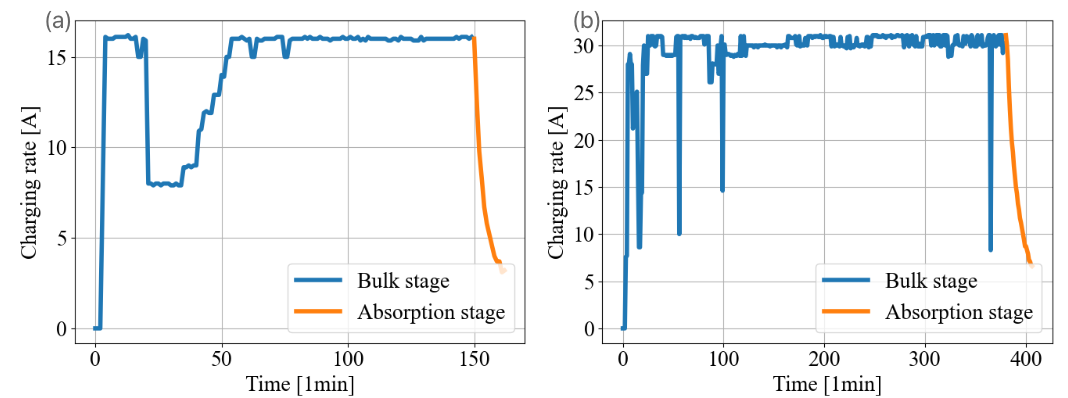}
\caption{Two-stage battery charging curves with distinct temporal patterns.}
\vspace{-10pt}
\label{charging curve}
\end{figure}

\subsection{Battery Charging Rate Curve Generation}
In this paper, we are interested in generating the fine-grained time-series for active charging sessions. Denote the charging curve as $\mathbf{r}\in\mathbb{R}^{L_{r}}$, and each point $\mathbf{r}_{i}$ at time $i,i=1,...,L_{r}$ refers to the exact charging rate drawn from the charging pile, and it can be either recorded in current or power based on available measurements. $L_{r}$ indicates the valid charging duration, which is a random variable due to the varying battery initial SOC, requested energy and charging circumstances. We wish the generated set of battery charging curves should \emph{involve different values of charging duration $L_{r}$}.

In general, charging curves of nowadays batteries are normally composed of two nonlinear stages, namely the bulk and absorption stage. In Fig. \ref{charging curve}, we display two samples of real battery charging curves. The sub-sequences of the bulk and absorption stage are rendered in blue and orange respectively, and the segmentation method proposed in~\cite{Chenxi2020} is adopted to get such divisions. In the bulk stage, charging curves exhibit a roughly steady trend with stochastic fluctuations. In the absorption stage, different decline forms can manifest unique electrochemical properties of batteries, which provide valuable information for quantitative analysis of battery physics. In this regard, we ought to \emph{derive such temporal tendency and variations of the bulk stage} and \emph{reveal the distinctive decline features of the absorption stage}. \textcolor{black}{Our goal is thus to accurately approximate the joint probability distribution $q(\mathbf{r})$ over battery charging rate time-series $\mathbf{r}$ based on their limited records,} and then generate a host of realistic charging sessions represented by various duration and distinct temporal dynamics to reflect individual EV charging properties.

\begin{figure}[t]
\centering
\setlength{\abovecaptionskip}{0.cm}
\includegraphics[width=0.48\textwidth]{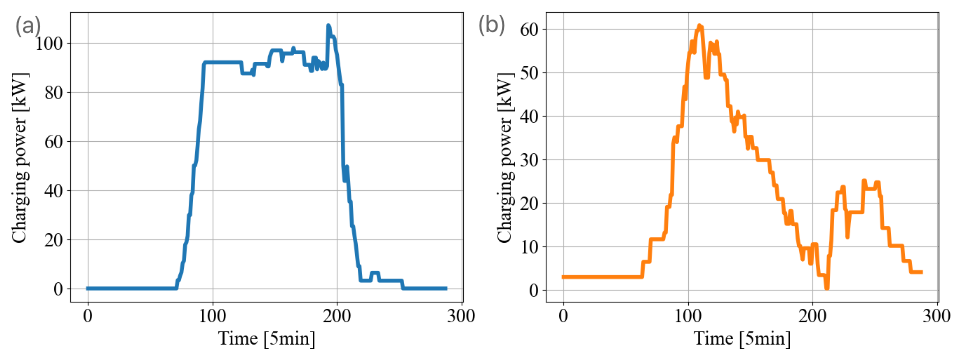}
\caption{Charging load profiles in different stations. (a) Workplace. (b) Campus.}
\vspace{-10pt}
\label{load profile}
\end{figure}

\subsection{Station Charging Load Profile Generation}
Obtaining reliable station-level charging load modes is also of potential interests for charging station and distribution grid operations. \textcolor{black}{Note that as station charging demand is stacked by total single charging sessions, we indeed can directly apply the battery-level diffusion model to produce an array of individual charging curves and aggregate them into overall load profiles. But this paradigm requires to estimate relationships between the number of daily charged EVs and their arrival time with generated duration. Besides, the charging capacity limit of different stations exerts another restriction on the aggregation manner. Since we aim to procure plausible demand patterns for different charging stations which are significant to system operations, customizing a separate diffusion model using real station-level charging scenarios could be the better choice.}

Let $\mathbf{s}\in\mathbb{R}^{L_{s}}$ denotes the charging load profile which aggregates all of daily EV charging sessions. Different temporal features of $\mathbf{s}$ can represent disparate station-level charging demands and usage patterns. Note that $L_{s}$ is the time horizon of one day, which is a constant and depends on the stipulated time interval. Every point $\mathbf{s}_{i}$ indicates the total charging power delivered to concurrent charged EVs at time $i,i=1,...,L_{s}$. Our goal on station-level generation is to yield realistic charging load profiles with unique temporal patterns for different charging stations. As shown in Fig. \ref{load profile}, the charging station at workplace displays distinguishable peak values and power variations from those on campus. Accordingly, we focus on learning the conditional joint distribution $q(\mathbf{s}|y_{s})$, i.e., given the label of a specific charging station $y_{s}$, we hope to produce a set of charging load time-series with consistent temporal modes belonging to that station.

\section{Denoising Diffusion Models}
\label{sec:pre}
\begin{figure*}[t]
\centering
\includegraphics[width=0.9\textwidth]{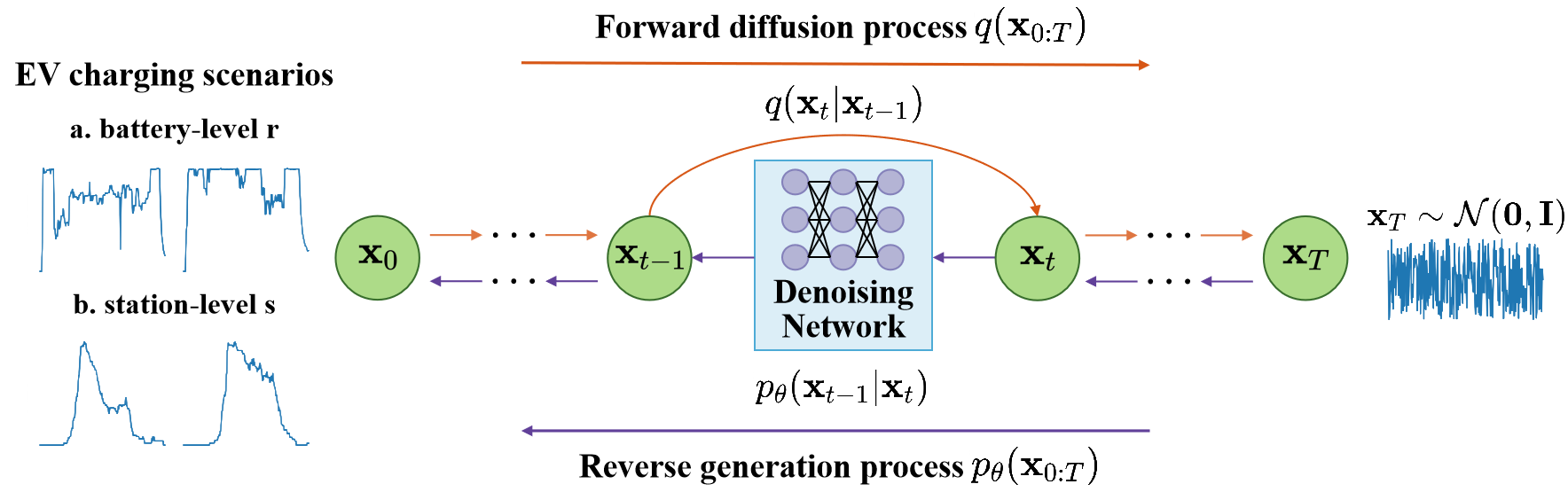}
\caption{Overview of the proposed DiffCharge framework based on the denoising diffusion model for EV charging scenario generation.}
\vspace{-10pt}
\label{framework}
\end{figure*}

In this section, we shed light on the principles of denoising diffusion models for scenario generation, which depend on learning to invert the diffusion process step-by-step to synthesize new charging time-series. We illustrate our DiffCharge framework which is grounded on DDPMs \cite{Ho2020} in Fig. \ref{framework}. Both training and sampling algorithms are distilled in a simple and straightforward form which are efficient to execute. \textcolor{black}{For ease of notation, we use $\mathbf{x}$ to denote charging time-series that will be harnessed in diffusion modeling, which indicates either battery-level curves $\mathbf{r}$ or station-level profiles $\mathbf{s}$.}

\subsection{Forward Diffusion Process}
Diffusion models have received stunning attention due to their strong capacity to generate high-dimensional complex data and better training efficiency compared to other generative models \cite{dhariwal2021diffusion}. We first exploit the forward diffusion to transform the real charging time-series data into Gaussian noise, and then the reverse diffusion to restore original data from the perturbed noise. Both forward and reverse procedures can be defined in form of Markov Chain, and the reverse Gaussian transitions can be learned by a deep denoising neural network \cite{Ho2020}. \textcolor{black}{Note that we use $\mathbf{x}_{0}$ to denote the original pure data (i.e. real $\mathbf{r}$ and $\mathbf{s}$), and $\mathbf{x}_{t}$ for the corrupted data at each diffusion step $t$.}

During the forward process, the step-wise Gaussian noise gradually degrades the raw charging time-series $\mathbf{x}_{0}$. After $T$ steps of diffusion, the real data distribution $q(\mathbf{x}_{0})$ is ultimately destroyed into a simple Gaussian distribution. This forward noise addition process can be defined as a fixed Markov Chain:
\begin{equation}
\setlength\abovedisplayskip{4pt}
\setlength\belowdisplayskip{4pt}
\label{eq:1}
q(\mathbf{x}_{1:T}|\mathbf{x}_{0})=\prod_{t=1}^{T}q(\mathbf{x}_{t}|\mathbf{x}_{t-1}),
\end{equation}
where $\mathbf{x}_{t},t=1,...,T$ can be deemed as latent variables, which reflect intermediate outcomes after real charging profiles are perturbed by $t$-step Gaussian noise. $q(\mathbf{x}_{t}|\mathbf{x}_{t-1})$ denotes the forward Markov transition, which reveals the mean and variance of the Gaussian noise added on $\mathbf{x}_{t-1}$:
\begin{equation}
\setlength\abovedisplayskip{4pt}
\setlength\belowdisplayskip{4pt}
\label{eq:2}
q(\mathbf{x}_{t}|\mathbf{x}_{t-1})=\mathcal{N}(\mathbf{x}_{t};\sqrt{1-\beta _{t}}\mathbf{x}_{t-1},\beta_{t}\mathbf{I}),
\end{equation}
where $\beta_{t}$ is the diffusion rate (i.e. the variance of Gaussian noise) at step $t$, which will escalate along with the forward process. The exact value of $\beta_{t}$ can be determined by the noise schedule scheme, which will be specified in the next section. After the $T$-step diffusion procedure, real charging scenarios are eventually transformed into a simple Gaussian $\mathbf{x}_{T}$, which is convenient to sample and manipulate.

A significant property regarding the forward diffusion is that we can explicitly calculate $\mathbf{x}_{t}$ at arbitrary intermediate steps based on the initial EV charging time-series $\mathbf{x}_{0}$ \cite{Ho2020}:
\begin{equation}
\label{eq:3}
\setlength\abovedisplayskip{4pt}
\setlength\belowdisplayskip{4pt}
q(\mathbf{x}_{t}|\mathbf{x}_{0})=\mathcal{N}(\mathbf{x}_{t};\sqrt{\alpha_{t}}\mathbf{x}_{0},(1-\alpha_{t})\mathbf{I}),
\end{equation}
where $\alpha_{t}=\prod_{s=1}^{t}(1-\beta_{s})$. Then, we can derive the closed-form expression of $\mathbf{x}_{t}$:
\begin{equation}
\setlength\abovedisplayskip{4pt}
\setlength\belowdisplayskip{4pt}
\label{eq:4}
\mathbf{x}_{t}=\sqrt{\alpha_{t}}\mathbf{x}_{0}+\sqrt{1-\alpha_{t}}\bm{\epsilon},\;
\bm{\epsilon}\sim\mathcal{N}(\mathbf{0},\mathbf{I}).
\end{equation}

Based on \eqref{eq:4}, $\mathbf{x}_{t}$ can be interpreted as the linear combination of $\mathbf{x}_{0}$ and the standard Gaussian $\bm{\epsilon}$. The concrete proof of \eqref{eq:3} and \eqref{eq:4} is evidenced in \cite{Calvin2022}, which adopts the reparameterization trick to rewrite $\mathbf{x}_{t}$ by $\mathbf{x}_{0}$ and $\bm{\epsilon}$.

\subsection{Reverse Generation Process}
To gain realistic charging time-series data, a progressive reverse process is designed, which can inversely convert the simple Gaussian to the real charging data distribution. \textcolor{black}{The motivation behind such denoising process is from the non-equilibrium thermodynamics~\cite{song2020score}}: if we can figure out the step-wise Gaussian noise injected in the forward process, we will restore the real charging time-series distribution through a series of iterative denoising steps. Accordingly, the reverse process starts from a pure Gaussian $\mathbf{x}_{T}$ and gradually removes the step-wise Gaussian noise from the perturbed $\mathbf{x}_{t}$, until the authentic charging scenarios are completely recovered. Such reverse process can be represented as:
\begin{equation}
\setlength\abovedisplayskip{4pt}
\setlength\belowdisplayskip{4pt}
\label{eq:5}
p_{\theta}(\mathbf{x}_{0:T})=p(\mathbf{x}_{T})\prod_{t=1}^{T} p_{\theta}(\mathbf{x}_{t-1}|\mathbf{x}_{t}),
\end{equation}
where $\mathbf{x}_{T}\sim\mathcal{N}(\mathbf{0},\mathbf{I})$ and $p_{\theta}$ indicates the reverse Markov transition. In light of the statistical properties of the continuous diffusion process, if the amount of added Gaussian noise (i.e. the noise variance $\beta$) is small enough, the reverse transition $p_{\theta}(\mathbf{x}_{t-1}|\mathbf{x}_{t})$ owns identical function form to forward transition $q(\mathbf{x}_{t}|\mathbf{x}_{t-1})$. Consequently, $p_{\theta}$ can be formed as a parameterized Gaussian distribution whose mean and variance can be approximated by a denoising neural network described later:
\begin{equation}
\setlength\abovedisplayskip{4pt}
\setlength\belowdisplayskip{4pt}
\label{eq:6}
p_{\theta}(\mathbf{x}_{t-1}|\mathbf{x}_{t})=\mathcal{N}(\mathbf{x}_{t-1}; \bm{\mu} _{\theta}(\mathbf{x}_{t},t),\mathbf{\Sigma}_{\theta}(\mathbf{x}_{t},t)),
\end{equation}
and our goal is to design an efficient way to learn to sample from $p_{\theta}(\mathbf{x}_{t-1}|\mathbf{x}_{t})$, so that real charging scenario distribution $q(\mathbf{x}_{0})$ can be ultimately derived by $p_{\theta}(\mathbf{x}_{0})$.

\subsection{Training Objective}
To solve the generation process \eqref{eq:6}, we approximate the joint distribution of real charging temporal data by maximizing their log-likelihoods. But the log-likelihood function $\log p_{\theta}(\mathbf{x}_{0})$ is intractable and hard to be computed explicitly. In this regard, we opt to maximize its evidence lower bound (ELBO) as an alternative to deal with it \cite{Ho2020}, which can be partitioned into a combination of closed-form expressions:
\begin{equation}
\setlength\abovedisplayskip{4pt}
\setlength\belowdisplayskip{4pt}
\label{eq:7}
\log p_{\theta}(\mathbf{x}_{0})\ge ELBO=\mathbb{E}_{q(\mathbf{x}_{1:T}|\mathbf{x}_{0})}[\log \frac{p_{\theta}(\mathbf{x}_{0:T})}{q(\mathbf{x}_{1:T}|\mathbf{x}_{0})}].
\end{equation}

Note that the ELBO in \eqref{eq:7} can be derived via Jensen's inequality \cite{Calvin2022}, and $q(\mathbf{x}_{1:T}|\mathbf{x}_{0})$ is the posterior of $p_{\theta}(\mathbf{x}_{0})$, which incorporates a series of latent variables $\mathbf{x}_{1:T}$ in the forward process and can be easily calculated by \eqref{eq:2}. Next, maximizing $\log p_{\theta}(\mathbf{x}_{0:T})$ can be equivalent to minimizing its negative ELBO, which can be decomposed into $T+1$ tractable items for explicit calculation:
\begin{equation}
\setlength\abovedisplayskip{4pt}
\setlength\belowdisplayskip{4pt}
\label{eq:8}
-ELBO=\mathcal{L}_{T}+\sum_{t=2}^{T}\mathcal{L}_{t-1}+\mathcal{L}_{0}.
\end{equation}

\textcolor{black}{In the following content, we shed light on how to obtain exact expressions of $\mathcal{L}_{T}$, $\mathcal{L}_{t-1}$ and $\mathcal{L}_{0}$ respectively for the EV charging scenario generation problem, and more detailed derivation can be found in the appendix of~\cite{Ho2020}}:
\begin{subequations}
\setlength\abovedisplayskip{4pt}
\setlength\belowdisplayskip{4pt}
\begin{align}
\label{eq:9a}
&\mathcal{L}_{T}=\mathcal{D}_{KL}(q(\mathbf{x}_{T}|\mathbf{x}_{0})||p_{\theta}(\mathbf{x}_{T}));\\
\label{eq:9b}
&\mathcal{L}_{t-1}=\mathbb{E}_{q(\mathbf{x}_{t}|\mathbf{x}_{0})}[\mathcal{D}_{KL}(q(\mathbf{x}_{t-1}|\mathbf{x}_{t},\mathbf{x}_{0})||p_{\theta}(\mathbf{x}_{t-1}|\mathbf{x}_{t}))];\\
\label{eq:9c}
&\mathcal{L}_{0}=-\mathbb{E}_{q(\mathbf{x}_{1}|\mathbf{x}_{0})}[\log p_{\theta}(\mathbf{x}_{0}|\mathbf{x}_{1})].
\end{align}
\end{subequations}

In \eqref{eq:9a}-\eqref{eq:9c}, we elaborate how to compute each item of \eqref{eq:8} in a closed-form manner. Initially, $\mathcal{L}_{T}$ indicates the distribution discrepancy between the last state of the forward process and the first state of the reverse process, both of which shall obey the standard Gaussian distribution. Hence, $\mathcal{L}_{T}$ is equal to zero and without trainable parameters. $\mathcal{L}_{0}$ is a reconstruction term which can be handled as a special case (i.e. $t=1$) of $\mathcal{L}_{t-1}$. When $t=1$, $\mathcal{L}_{t-1}$ amounts to minimizing the KL divergence between $q(\mathbf{x}_{0})$ and $p_{\theta}(\mathbf{x}_{0}|\mathbf{x}_{1})$, which holds the consistent training target with $\mathcal{L}_{0}$ stated in \eqref{eq:9c}. Now we only need to figure out $\mathcal{L}_{t-1}$, which can be interpreted as the supervised noise elimination. In specific, $p_{\theta}(\mathbf{x}_{t-1}|\mathbf{x}_{t})$ indicates the Gaussian noise added at step $t$ is removed from the perturbed $\mathbf{x}_{t}$ by the denoising network, in other words, $\mathbf{x}_{t}$ is denoised into $\mathbf{x}_{t-1}$. $q(\mathbf{x}_{t-1}|\mathbf{x}_{t},\mathbf{x}_{0})$ reflects the ground-truth signal for the output of the denoising network. Note that $q(\mathbf{x}_{t-1}|\mathbf{x}_{t},\mathbf{x}_{0})$ can be analytically calculated as follows:
\begin{align}
\label{eq:10}
q(\mathbf{x}_{t-1}|\mathbf{x}_{t},\mathbf{x}_{0})
& = \frac{q(\mathbf{x}_{t}|\mathbf{x}_{t-1},\mathbf{x}_{0})q(\mathbf{x}_{t-1}|\mathbf{x}_{0})}{q(\mathbf{x}_{t}|\mathbf{x}_{0})} \nonumber \\
& = \mathfrak{\mathcal{N}} (\mathbf{x}_{t-1};\tilde{\bm{\mu}} _{t}(\mathbf{x}_{t},\mathbf{x}_{0}),\tilde{\beta} _{t}\mathbf{I}).
\end{align}

The Bayesian rule is applied in the first line of \eqref{eq:10}, and the conditional probability is tractable as the right hand side of \eqref{eq:10} are three known  Gaussian distributions. The mean and variance defined in \eqref{eq:10} are presented as follows \cite{Ho2020}:
\begin{align}
\setlength\abovedisplayskip{4pt}
\setlength\belowdisplayskip{4pt}
\label{eq:11}
\tilde{\bm{\mu}}_{t}(\mathbf{x}_{t},\mathbf{x}_{0})
& = \frac{\sqrt{\alpha_{t-1}}\beta_{t}}{1-\alpha_{t}}\mathbf{x}_{0}+\frac{\sqrt{1-\beta_{t}}(1-\alpha_{t-1})}{1-\alpha_{t}}\mathbf{x}_{t} \nonumber \\
& = \frac{1}{\sqrt{1-\beta_{t}}}(\mathbf{x}_{t}-\frac{\beta_{t}}{\sqrt{1-\alpha_{t}}}\bm{\epsilon});\\
\tilde{\beta}_{t}&=\frac{1-\alpha_{t-1}}{1-\alpha_{t}}{\beta}_{t}.
\end{align}
Note that in \eqref{eq:10}, $\mathbf{x}_{0}$ can be derived by $\mathbf{x}_{t}$ and $\bm{\epsilon}$ based on \eqref{eq:4}. Similar with the empirical simplification adopted in \cite{Ho2020}, we also fix $\bm{\Sigma}_{\theta}(\mathbf{x}_{t},t) =\tilde{\beta}_{t}\mathbf{I}$ for $p_{\theta}(\mathbf{x}_{t-1}|\mathbf{x}_{t})$, which circumvents the burden to learn the variance of the reverse transition. Therefore, $\mathcal{L}_{t-1}$ can be reduced to calculate the KL divergence between two Gaussian distributions defined in \eqref{eq:6} and \eqref{eq:10}, which have different means yet the same variance. Then $\mathcal{L}_{t-1}$ can be computed as $\mathcal{L}_{t-1}=\mathbb{E}_{q(\mathbf{x}_{t}|\mathbf{x}_{0})}[\frac{1}{2\tilde {\beta}_{t}}\left\|\tilde{\bm{\mu}}_{t}(\mathbf{x}_{t},\mathbf{x}_{0})-\bm{\mu}_{\theta}(\mathbf{x}_{t},t)\right\|_2^{2}]$ \cite{Ho2020},
which is to predict the mean of the reverse transition. 
However, based on \eqref{eq:11} and the reparameterization technique proposed in \cite{Ho2020}, \textcolor{black}{we opt the form for $\bm{\mu} _{\theta}(\mathbf{x}_{t},t)$ as follows}:
\begin{equation}
\label{eq:13}
\setlength\abovedisplayskip{4pt}
\setlength\belowdisplayskip{4pt}
\bm{\mu} _{\theta}(\mathbf{x}_{t},t)=\frac{1}{\sqrt{1-\beta_{t}}}(\mathbf{x}_{t}-\frac{\beta_{t}}{\sqrt{1-\alpha_{t}}}\bm{\epsilon}_{\theta }(\mathbf{x}_{t},t)),
\end{equation}
which \textcolor{black}{points out} a more efficient way to train the denoising network, i.e. directly predicting the Gaussian noise $\bm{\epsilon}$ added at step $t$ instead of the mean $\tilde\mu_{t}$ of the reverse transition. To this end, \textcolor{black}{calculation of $\mathcal{L}_{t-1}$ can be adapted to a more straightforward form }\cite{Ho2020}:
\begin{equation}
\setlength\abovedisplayskip{4pt}
\setlength\belowdisplayskip{4pt}
\label{eq:14}
\mathcal{L}_{t-1}=\mathbb{E}_{\mathbf{x}_{0},\bm{\epsilon},t}[\left\| \bm{\epsilon}-\bm{\epsilon}_{\theta}(\sqrt{\alpha_{t}}\mathbf{x}_{0}+\sqrt{1-\alpha_{t}} \bm{\epsilon},t)\right\|_2^{2}],
\end{equation}
where $t$ can be uniformly sampled from $[1, T]$, and $\bm{\epsilon}_{\theta}(\cdot )$ represents the denoising neural network whose input is the perturbed $\mathbf{x}_{t}$ defined in \eqref{eq:4} and output is the prediction for the $t$-step Gaussian noise $\bm{\epsilon}$. Once the denoising network $\bm{\epsilon}_{\theta}(\cdot )$ is well trained using \eqref{eq:14}, we can utilize it to gradually generate plausible charging scenarios along with the reverse denoising process by finding $ \bm{\mu} _{\theta}(\mathbf{x}_{t},t)$ based on \eqref{eq:13}. The coordinated training and sampling algorithms for the denoising diffusion models are presented in Algorithm \ref{algo1}. \textcolor{black}{We stop the training procedure when the training loss defined in \eqref{eq:14} ceases to decrease for a predefined number of epochs.}

\begin{algorithm}
  \caption{Training and Sampling of DDPMs}
  \label{algo1}
  \begin{algorithmic}
  \REQUIRE EV charging time-series data $q(\mathbf{x}_{0})$ \\
    \# \textit{Training stage}
    \WHILE{$\theta$ has not converged}
    \STATE $\mathbf{x}_{0}\sim q(\mathbf{x}_{0})$, $\bm{\epsilon}\sim \mathcal{N}(\mathbf{0},\mathbf{I})$, $t\sim \text{Uniform}(\{1,...,T\})$
    \STATE Update the denoising network using gradient descent: \\
    $\bigtriangledown _{\theta }\left\| \bm{\epsilon}-\bm{\epsilon}_{\theta}(\sqrt{\alpha_{t}}\mathbf{x}_{0}+\sqrt{1-\alpha_{t}} \bm{\epsilon},t)\right\|_2^{2}$ using \eqref{eq:14}\\
    \ENDWHILE \\
    \# \textit{Sampling stage}
    \STATE $\mathbf{x}_{T}\sim \mathcal{N}(\mathbf{0},\mathbf{I})$ 
    \FOR{$t=T,...,1$}
    \STATE $\mathbf{z}\sim \mathcal{N}(\mathbf{0},\mathbf{I})$
    \STATE $\mathbf{x}_{t-1}=\frac{1}{\sqrt{1-\beta_{t}}}(\mathbf{x}_{t}-\frac{\beta_{t}}{\sqrt{1-\alpha_{t}}}\epsilon_{\theta}(\mathbf{x}_{t},t))+\sqrt{\tilde{\beta}_{t}}\mathbf{z}$ \\
    according to both \eqref{eq:6} and \eqref{eq:13}
    \ENDFOR
  \end{algorithmic}
\end{algorithm}

\section{Detailed designs for DiffCharge}
\label{sec:model}
As stated in \cite{Lee2019}, different factors such as initial SOC, battery types, station locations, exogenous pilot signals delivered by online controllers and random noises, can affect temporal patterns of EV charging sessions. Thus it is necessary for the developed generative model to preserve unique temporal features in both battery-level and station-level charging time-series. In this section, we show how to fit the denoising diffusion model introduced in the former section to attain EV charging scenario generation by employing attention mechanism, adapting noise schedule and enabling conditional generation.

\begin{figure}[t]
\centering
\setlength{\abovecaptionskip}{0.cm}
\includegraphics[width=0.5\textwidth]{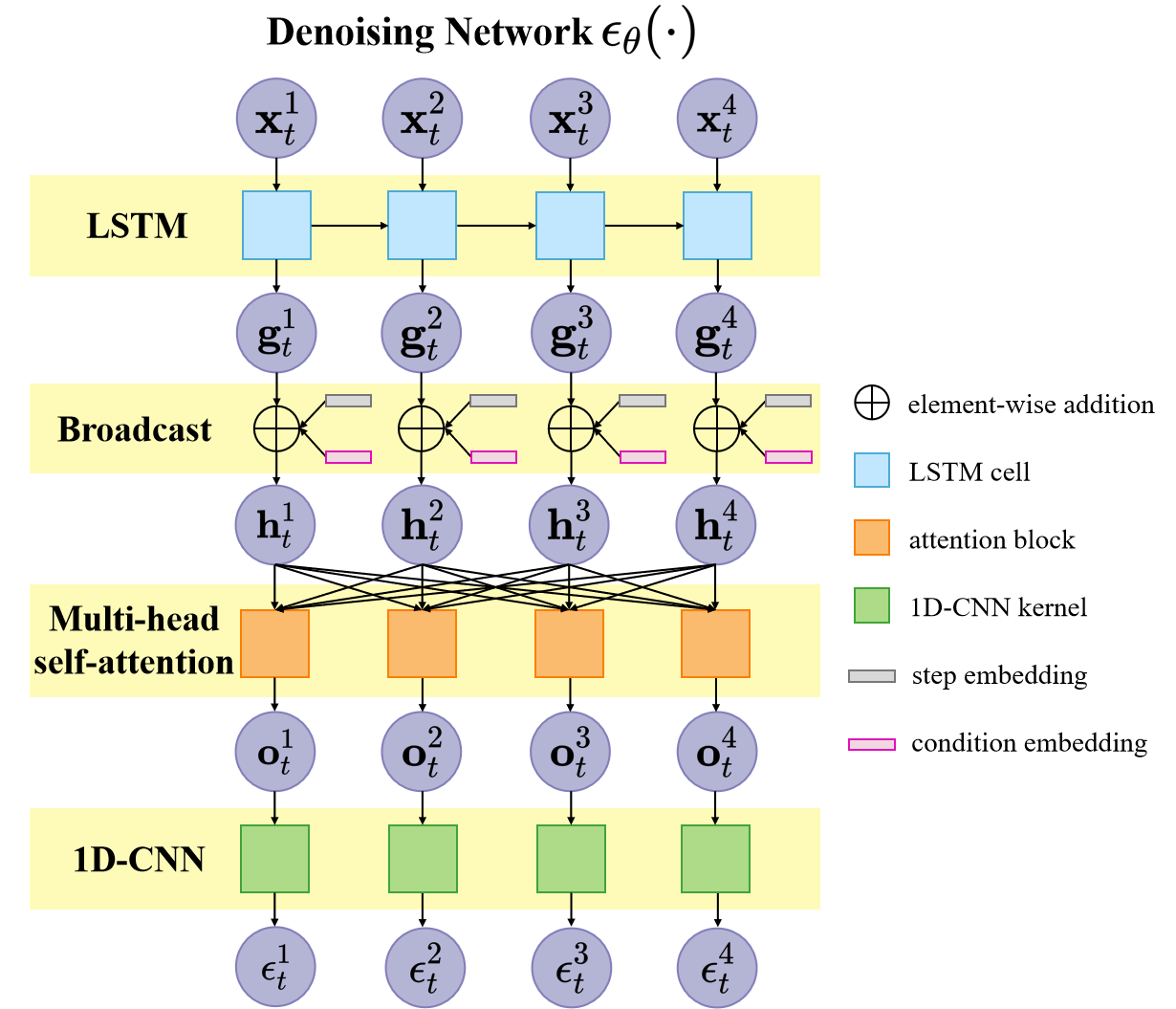}
\caption{The architecture diagram of the denoising network $\bm{\epsilon}_{\theta}(\cdot)$. For a more clear display, we set the length of time steps $L$ in $\mathbf{x}_{t}$ as 4 in this graph.}
\label{network}
\end{figure}

\subsection{Denoising Network}
Temporal correlations are ubiquitous in real-world charging load profiles. On the one hand, online scheduling strategies for individual EV charging usually involves the rolling optimization, which could take previous charging power into consideration for better user demand satisfaction. On the other hand, based on the electrochemical properties of battery charging dynamics \cite{Chen2022}, charging efficiencies are associated with the real-time SOC level, which are decided by the initial SOC and the integral of previous charging rates. Accordingly, we aim to design an adequate architecture for the denoising network $\bm{\epsilon}_{\theta}$, which is capable of preserving such complex temporal relations in the reverse generation process. 

\textcolor{black}{Under such design considerations, we integrate both long short-term memory (LSTM) network
and multi-head self-attention mechanism into the denoising network.} The combined LSTM and multi-head self-attention are adept at capturing temporal correlations in time-series data \cite{wen2023}. The overall architecture of the denoising network is illustrated in Fig. \ref{network}. We first utilize LSTM to get the latent state $\mathbf{g}_{t}\in \mathbb{R}^{L\times H}$ of the perturbed $\mathbf{x}_{t}\in \mathbb{R}^{L}$, where $L$ is the total number of time steps, $H$ is the dimension of hidden state. Note that for battery-level data $\mathbf{r}$, $L=L_{r}$, while for station-level data $\mathbf{s}$, $L=L_{s}$. At each diffusion step $t$, the noised $\mathbf{x}_{t}$ keeps the same length $L$ as original $\mathbf{x}_{0}$. We use $\mathbf{x}^{i}_{t}$ to denote the single value of $\mathbf{x}_{t}$ at each time $i$. For one LSTM cell $LSTM(\cdot)$, the hidden state $\mathbf{g}^{i}_{t}\in \mathbb{R}^{H}$ at time $i$ is calculated using both previous state $\mathbf{g}^{i-1}_{t}$ and the current observation $\mathbf{x}^{i}_{t}$. $\mathbf{g}^{i}_{t}$ can be calculated as $\mathbf{g}^{i}_{t}=LSTM(\mathbf{x}^{i}_{t}, \mathbf{g}^{i-1}_{t})$ \cite{Li2023}.

Then, we integrate the hidden state $\mathbf{g}^{i}_{t}$, condition embedding $\mathbf{c}$ and diffusion embedding $\mathbf{t}_{e}$ in the broadcast layer. In \eqref{eq:14}, as the denoising network should take each diffusion step $t$ as additional input, we design a diffusion-step embedding scheme to represent $t$ and will elucidate it later. The condition embedding $\mathbf{c}$ is applied to realize the conditional scenario generation, which will be further introduced at the end of this section. In this layer, we broadcast both $\mathbf{c}$ and $\mathbf{t}_{e}$ over each time $i$ to get latent encoding $\mathbf{h}^{i}_{t}\in \mathbb{R}^{H}$ (in practice, repeat $\mathbf{c}$ and $\mathbf{t}_{e}$ for $L$ times), and integrate them with the corresponding hidden state using  element-wise addition operation $\oplus$:
\begin{equation}
\setlength\abovedisplayskip{4pt}
\setlength\belowdisplayskip{4pt}
\label{eq:15}
\mathbf{h}^{i}_{t}= \mathbf{g}^{i}_{t}\oplus \mathbf{c}\oplus \mathbf{t}_{e}.
\end{equation}

Next, we introduce the multi-head self-attention mechanism embodied in Transformer \cite{wen2023}, which exhibits exceptional capacity to extract significant information from sequential data. Multi-head attention is employed in our denoising network to sufficiently represent the complex correlations among all fused features $\mathbf{h}^{i}_{t}$, where $\mathbf{h}^{i}_{t}$ at each time $i$ is related to the global $\mathbf{h}^{1:L}_{t}$. Let us clarify how multi-head attention takes effect. Firstly, the latent encoding $\mathbf{h}_{t}\in \mathbb{R}^{L\times H}$ are linearly projected into the query, key and value matrices in a single attention head using $\mathbf{Q}_{b}=\mathbf{h}_{t}\mathbf{W}^{Q}_{b},\mathbf{K}_{b}=\mathbf{h}_{t} \mathbf{W}^{K}_{b},\mathbf{V}_{b}=\mathbf{h}_{t}\mathbf{W}^{V}_{b}$,
where $b=1,...,B$ is the index of the attention head, $B$ is the number of attention heads. $\mathbf{W}^{Q}_{b},\mathbf{W}^{K}_{b},\mathbf{W}^{V}_{b}\in \mathbb{R}^{H\times d_{b}}$ are learnable parameters and $d_{b}$ is the feature dimension of three matrices. Then, the scaled dot-product attention is adopted to calculate the output matrix $\mathbf{A}_{b}$ of the $b$-th attention head \cite{wen2023}:
\begin{equation}
\setlength\abovedisplayskip{4pt}
\setlength\belowdisplayskip{4pt}
\label{eq:16}
\mathbf{A}_{b}=softmax(\frac{\mathbf{Q}_{b}\mathbf{K}^{T}_{b}}{\sqrt{d_{b}}}) \mathbf{V}_{b}.
\end{equation}

Then the output matrices of $B$ heads $\mathbf{A}_{b}\in \mathbb{R}^{L\times d_{b}}$ are concatenated and processed by the linear transformation to realize the ultimate outcomes of multi-head attention:
\begin{equation}
\setlength\abovedisplayskip{4pt}
\setlength\belowdisplayskip{4pt}
\label{eq:17}
\mathbf{O}_{t}=[\mathbf{A}_{1},...,\mathbf{A}_{B}]\mathbf{W}^{O},
\end{equation}
where $\mathbf{W}^{O}\in \mathbb{R}^{D\times D}$ need to be learned, and $D=Bd_{b}$ is the feature dimension of the multi-head attention output. In Fig.~\ref{network}, the attention block indicates the partly calculation in multi-head attention which accounts for the single output $\mathbf{O}^{i}_{t}$ at each time $i$, and illustrates that $\mathbf{O}^{i}_{t}$ attach holistic attention to $\mathbf{h}^{i}_{t}$ at all time steps. At time $i$ for diffusion step $t$, the final output is $\bm{\epsilon}^{i}_{t}=CNN(\mathbf{O}^{i}_{t}, \mathbf{W}^{C}, \mathbf{b}^{bias})$, where $\mathbf{W}^{C}\in \mathbb{R}^{D}$ and $\mathbf{b}^{bias}$ are parameters of the 1D-CNN layer. 

In order to make the trained neural network $\bm{\epsilon}_{\theta}$ perceive different values of diffusion step $t$, we develop a diffusion-step embedding scheme, which is consistent with the positional encoding in \cite{wen2023}. The sine and cosine functions are utilized for this step embedding. The dimension of such embedding vectors are particularly set to $H$, which is equal to the hidden state in LSTM. For $j=0,1,...,\frac{H}{2}-1$, the proposed embedding scheme can be exhibited as:
\begin{subequations}
\setlength\abovedisplayskip{4pt}
\setlength\belowdisplayskip{4pt}
\begin{align}
\label{eq:18a}
& t_{e}(2j)=sin(t/10000^{2j/H}), \\
\label{eq:18b}
& t_{e}(2j+1)=cos(t/10000^{2j/H}).
\end{align}
\end{subequations}

\subsection{Noise Schedule}
In light of recent \textcolor{black}{advancements} on DDPMs \cite{Nichol2021}, selecting a proper variance $\beta_{t}$ of Gaussian noise at every diffusion step is important to achieve the desired high-quality generation. Since the denoising network aims to accurately predict the noise added on the perturbed $\mathbf{x}_{t}$, well-scheduled noise that transfers raw data $\mathbf{x}_{0}$ to $\mathbf{x}_{t}$ is the key to successful model training. To this end, we leverage the quadratic schedule involved in \cite{tashiro2021csdi}, which can smoothly add noise step-by-step and be beneficial for the generative ability of diffusion models:
\begin{equation}
\setlength\abovedisplayskip{4pt}
\setlength\belowdisplayskip{4pt}
\label{eq:19}
\beta_{t}=(\frac{T-t}{T-1}\sqrt{\beta_{1}}+\frac{t-1}{T-1}\sqrt{\beta_{T}})^{2}.
\end{equation}

\begin{figure}[t]
\centering
\setlength{\abovecaptionskip}{0.cm}
\includegraphics[width=0.4\textwidth]{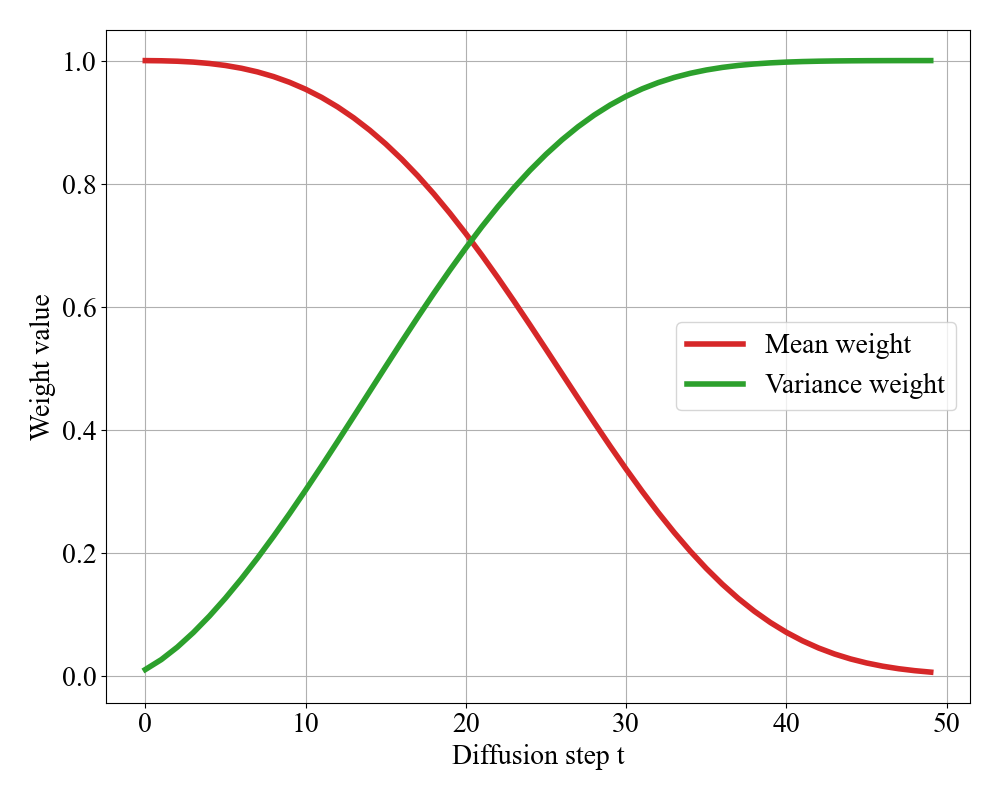}
\caption{Weights $\sqrt{\alpha_{t}}, \sqrt{1-\alpha_{t}}$ associated with mean and variance term presented in \eqref{eq:4} under the quadratic noise schedule.}
\label{noise}
\end{figure}

Here we follow the common setting that $\beta_{1}=0.0001$ and $\beta_{T}=0.5$ \cite{tashiro2021csdi}. The number of diffusion steps $T$ is also a critical parameter, as larger $T$ can render the reverse transition defined in \eqref{eq:6} much closer to the Gaussian form. For image generation, it always requires thousands of diffusion steps, which sacrifice the sampling speed. But in our experiments, since charging profiles are relatively less intricate than images, we find fixing $T=50$ is adequate to procure satisfactory outcomes on our double-level charging time-series generation. After $\beta_{t}$ is ascertained, in Fig. \ref{noise}, we plot the weight $\sqrt{\alpha_{t}}, \sqrt{1-\alpha_{t}}$ associated with mean and variance terms in \eqref{eq:4} for our simulation. We can observe that with the increase of diffusion step $t$, $\sqrt{\alpha_{t}}$ descends whereas $\sqrt{1-\alpha_{t}}$ ascends. The weight of real sample $\mathbf{x}_{0}$ is gradually diminishing while noise weight  $\bm{\epsilon}$ is steadily increasing, meaning $\mathbf{x}_{0}$ is progressively perturbed by noise $\bm{\epsilon}$ in forward diffusion process.

\subsection{Conditional Scenario Generation}
As mentioned in Section \ref{sec:formulation}, in practical EV charging operation, it is preferred to generate diverse charging load profiles for different station types, as their daily charging demands and usage patterns can be quite unique. To this end, we need to upgrade our unconditional diffusion model discussed above to a conditional one. And the major goal is to approximate the conditional joint distribution $q(\mathbf{s}|y_{s})$, where $y_{s}$ denotes a specific charging station. We can straightforwardly attain the diffusion-based conditional synthesis by modifying the reverse process defined in \eqref{eq:5} and \eqref{eq:6} to a conditional version:
\begin{subequations}
\setlength\abovedisplayskip{4pt}
\setlength\belowdisplayskip{4pt}
\begin{align}
\label{eq:20}
&p_{\theta}(\mathbf{x}_{0:T}|y_{s})=p(\mathbf{x}_{T})\prod_{t=1}^{T} p_{\theta}(\mathbf{x}_{t-1}|\mathbf{x}_{t},y_{s});\\
&p_{\theta}(\mathbf{x}_{t-1}|\mathbf{x}_{t},y_{s})=\mathcal{N}(\mathbf{x}_{t-1}; \bm{\mu}_{\theta}(\mathbf{x}_{t},y_{s},t),\tilde{\beta}_{t}\mathbf{I}).
\end{align}
\end{subequations}

Thereby, the training objective $\mathcal{L}_{t-1}$ defined in \eqref{eq:14} naturally becomes the following form:
\begin{equation}
\setlength\abovedisplayskip{4pt}
\setlength\belowdisplayskip{4pt}
\label{eq:21}
\mathcal{L}^{cond}_{t-1} = \mathbb{E}_{\mathbf{x}_{0},y_{s},\bm{\epsilon},t}[\left\|\bm{\epsilon}-\bm{\epsilon}_{\theta}(\sqrt{\alpha_{t}}\mathbf{x}_{0}+\sqrt{1-\alpha_{t}} \bm{\epsilon},y_{s},t)\right\|_2^{2}].
\end{equation}

\textcolor{black}{Consequently, the denoising network $\bm{\epsilon}_{\theta}$ should take the embedding $\mathbf{c}$ of the discrete condition term $y_{s}$ as additional input. Such condition input in \eqref{eq:15} and the related broadcast layer are illustrated in Fig. \ref{network}.}

\section{Experimental Setup}
\label{sec:setup}
\subsection{Dataset Descriptions and Experiment Setup}
We employ ACN-Data \cite{Lee2019} which contains abundant real-world charging sessions of individual EVs collected in California. Among lots of detailed session information, we focus on a set of discrete attributes: connectionTime, doneChargingTime, kWhDelivered and fine-grained charging current signals (indicating charging rate curves). We also adopt the accumulation technique proposed in \cite{Li2023} to extract daily charging load profiles by aggregating overall charging sessions of daily served EVs at the station, which integrates various arrival/departure time and actual scheduled energy. In particular, we use session data collected at JPL and Caltech in 2018, which deposit sufficient charging demand profiles with distinct battery dynamics. As the JPL and Caltech station belong to workplace and campus respectively, they can exhibit disparate energy requirement and usage patterns. In practice, we down-sample original charging current curves to 1-min resolution, and fix the length of each curve to 12 hours by zero padding, i.e. setting $L_{r}=720$. Besides, we fetch daily station charging load profiles on a 5-min basis with $L_{s}=288$. For both synthesis tasks, we totally employ 1000 training samples.

We design a unified architecture for the denoising neural network in DiffCharge to learn two-fold charging scenarios, where the only disparity is the handled time-series lengths by the whole network (caused by different $L_{r}$ and $L_{s}$). The dimension of both the hidden state in the first LSTM layer and the latent embedding in the broadcast layer embedding are 48. We allocate 4 heads of self-attention blocks and also set their latent dimensions to 48 in the transformer encoder. Note that we simply unify the size of all hidden encodings, which saves our effort to adjust hyper-parameters. In the last 1D-CNN output layer, to make the predicted noise vector own the same size as the input degraded $\mathbf{x}_t$, its kernel size and stride are set to 48 and 1 respectively. DiffCharge is implemented by Pytorch and optimized by Adam with initial learning rate of 0.001. The training process is completed on a Linux service machine with a Nvidia 1080 Ti GPU, with 200 training epochs and a batch size of 4.

\textcolor{black}{\subsection{Baseline Models}
We compare our denoising diffusion-based generative model DiffCharge with three existing methods which also showcase impressive performance on time-series scenario generation.}

\textcolor{black}{GMM \cite{POWELL2022123592} is a parametric model-based generative method, which assumes charging demands (i.e. battery charging current or station charging power) over the whole time horizon comply to a mixed Gaussian distribution and strives to optimize its unknown parameters on the historical data. We manually set the mixed number of independent Gaussian components as 15.}

\textcolor{black}{VAEGAN \cite{de2021unsupervised} is based on the adversarial autoencoder and combines the advantage of VAEs and GANs. Its  encoder and decoder are formed by hybrid 1D-CNN and LSTM layers to handle intrinsic temporal features. We utilize its decoder to map the prior multivariate Gaussian to target scenarios.}

\textcolor{black}{TimeGAN \cite{yoon2019time} is an ad-hoc time-series generation model which integrates the autoregressive modeling to preserve the original temporal dynamics and adversarial training in GANs to capture the complex distribution of high-dimensional temporal sequences. We directly apply it on both battery-level and station-level charging time-series data.}

\begin{figure*}[t]
\setlength{\abovecaptionskip}{0.cm}
\centering
\includegraphics[width=0.8\textwidth]{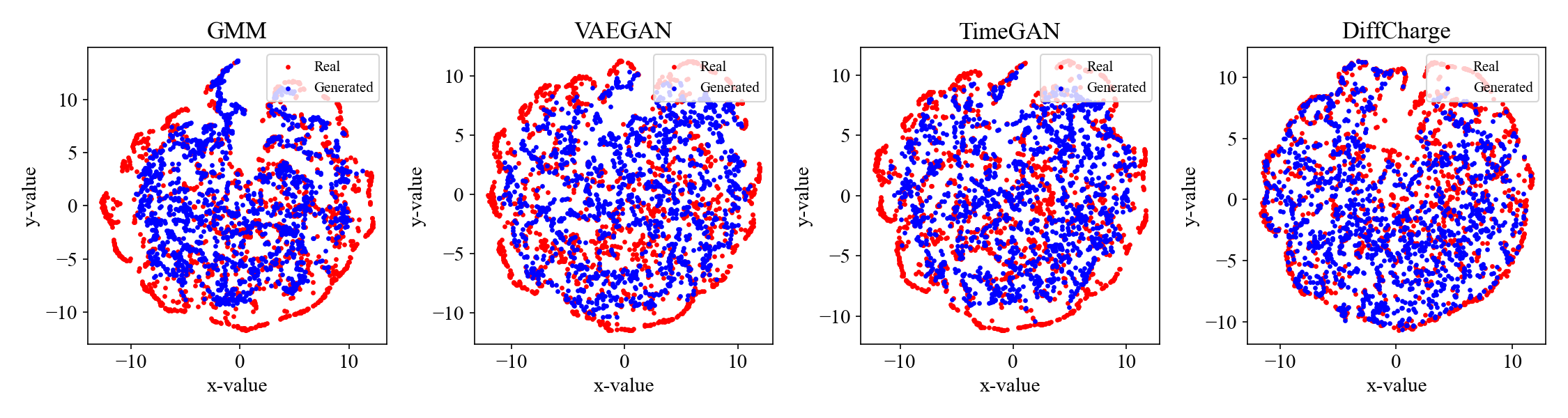}
\caption{\textcolor{black}{2D t-SNE visualizations of real and generated battery charging curves from different models.}}
\label{t-sne}
\end{figure*}

\textcolor{black}{\subsection{Evaluation Methods}
It is tough to explicitly calculate the high-dimensional distribution of generated charging loads and directly compare it with the ground-truth one. How to holistically assess their quality (i.e. the so-called realism, which reflects the distributional similarity) is of great importance. Considering both two-level charging scenarios are temporal data, we intend to incorporate the quantitative metrics proposed for generic time-series generation \cite{yoon2019time, zhou2023deep} as well as several unique perspectives tailored for charging curves \cite{Chenxi2020} and an application case \cite{jahangir2021novel} to inspect the generative performance of DiffCharge. Overall, there are three aspects to evaluate the generation quality: 1) \textit{diversity}, synthetic profiles are supposed to cover the complex temporal dynamics and various load patterns in original charging time-series; 2) \textit{fidelity}, generated samples shall be unidentifiable from real observations on temporal properties; 3) \textit{usefulness}, generated scenarios should bring some benefits to the practical EV charging integration. The evaluative criteria utilized in our experiments can encapsulate such three views.}

\textcolor{black}{We shed light on the evaluation methods for in our work. To begin with, we introduce three metrics for general time-series synthesis \cite{yoon2019time, zhou2023deep}: 1) \textit{Marginal score}, which calculates the absolute difference between two empirical marginal distributions (i.e. probability density function (PDF)) over charging demand values along all time steps. The lower score indicates the better quality. 2) \textit{Discriminative score}. We train a post-hoc binary classifier (consists of two LSTM layers and a Linear projector) using the cross-entropy loss to recognize between real and generated charging profiles. Ideally, if the generated samples are realistic enough, the classifier can not distinguish them from original data and manifest a higher cross-entropy (the perfect case is 0.7). We unravel the binary cross-entropy on the test set (containing 20\% real and generated samples) to reveal the synthesis quality. 3) \textit{2D visualization}. We adopt the t-SNE technique to visually analyze the distribution similarity in the two-dimensional space. It can depict the distribution of data samples to showcase the resemblance and coverage of generated temporal patterns in contrast to real situations.}

\textcolor{black}{Apart from these generic indicators, we also design domain-based standards to account for two special properties in battery charging sessions, i.e. valid duration and tail features in the absorption stage \cite{jahangir2021novel}. As for charging duration, we simply analyze its PDF to check whether DiffCharge can produce various lengths of charging curves (i.e. different values of $L_{r}$). With respect to tail features, we design the \textit{tail score} to assess the quality of generated temporal modes for the absorption stage. Similar to the clustering-based scoring method used in \cite{jahangir2021novel}, we divide the tail features of real curves into several clusters and predict which cluster each synthetic feature should lie in. For each cluster, we count two marginal cumulative density functions (CDFs) over charging rates in the tail feature for both real and generated samples respectively, and then compute the $L_{1}$-distance between real and generated CDF. Accordingly, the tail score can be deemed as the average CDF distance over all clusters. In Section \ref{sec:results}-D, we also provide an EV operation case in the day-head market to illustrate the usefulness of generated charging scenarios.}

\section{Results and Analysis}
\label{sec:results}
\textcolor{black}{In this section, we verify the effectiveness of DiffCharge based on real-world ACN-data and aforementioned evaluative methods. We conduct extensive experiments to corroborate the generation quality of both battery-level and station-level charging demand time-series using a set of quantitative criteria. In addition, we also build a practical operational case in the context of real-world electricity market to validate the utility of synthetic charging scenarios.}

\begin{figure*}[!t]
\centering
\setlength{\abovecaptionskip}{0.cm}
\includegraphics[width=\textwidth]{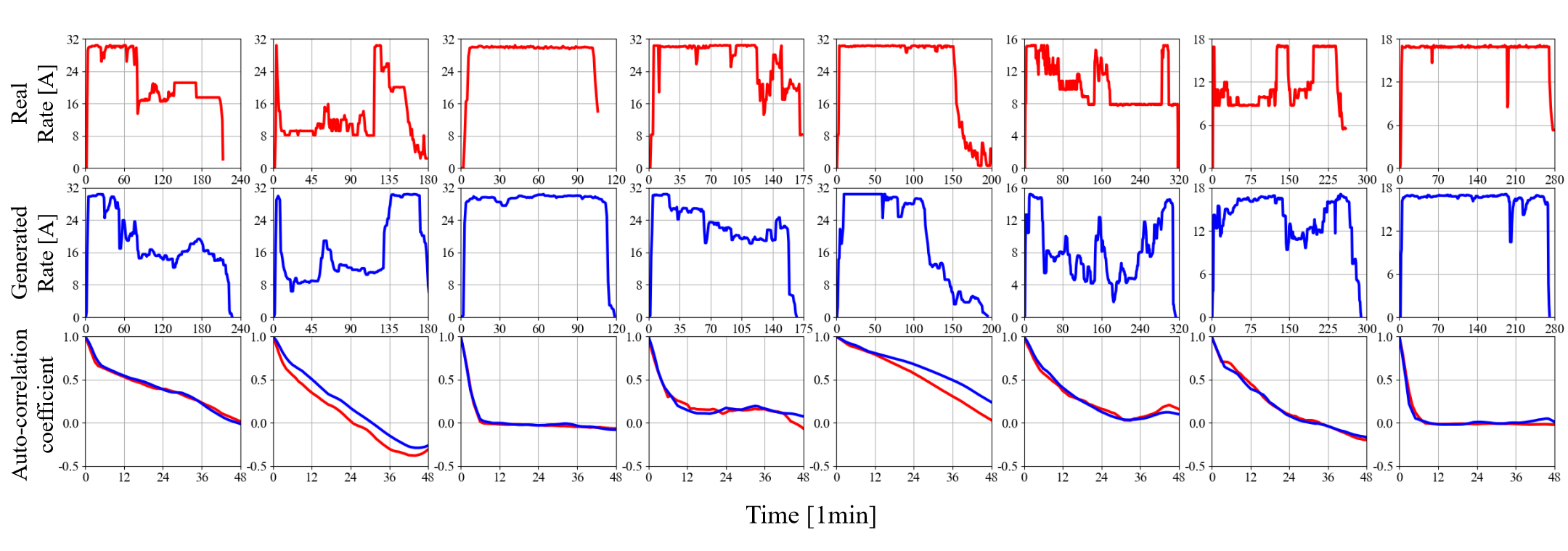}
\caption{Randomly selected real and generated battery charging curves, and their corresponding auto-correlation coefficients. Note that the x axis in the third row indicates the lag time when calculating auto-correlation coefficients for each charging curve, whose time unit is also 1 min.}
\label{acf}

\end{figure*}

\begin{table*}[!t]
\begin{center}
\caption{\textcolor{black}{Quantitative evaluation metrics results on 
battery-level generation.}}
\setlength{\abovecaptionskip}{0.cm}
\label{overall metrics}
\begin{tabular}{ccccc}
\hline\\[-2.9mm]\hline
Model         & Marginal Score ($\downarrow$) & Discriminative Score ($\uparrow$)  & Tail Score ($\downarrow$)            & \multicolumn{1}{l}{Inference Time {[}s{]}} \\ \hline
GMM           & 0.5705          & 0.4774±0.1098          & 0.1137±0.0631          & /                                          \\
VAEGAN        & 0.0827          & 0.5840±0.0742          & 0.0577±0.0384          & /                                          \\
TimeGAN       & 0.0782          & 0.5125±0.1991          & 0.0618±0.0407          & /                                          \\ \hline
w/o attention & 0.2339          & 0.2532±0.2071          & 0.0620±0.0228          & 159.8946                                   \\
$T=30$          & 0.1014          & 0.6195±0.0073          & 0.0657±0.0461          & 150.4856                                   \\
$T=40$          & 0.1011          & 0.6355±0.0155          & 0.0626±0.0460          & 190.1435                                   \\
$T=60$          & 0.1149          & 0.6398±0.0311          & 0.0633±0.0459          & 340.9384                                   \\
$T=70$          & 0.1088          & 0.5603±0.0265          & 0.0651±0.0447          & 388.2945                                   \\
DiffCharge    & \textbf{0.0534} & \textbf{0.6456±0.0180} & \textbf{0.0469±0.0144} & 251.0545                                   \\ 
\hline\\[-2.9mm]\hline
\end{tabular}
\begin{tablenotes}
\textcolor{black}{\item $\downarrow$ indicates lower score reflects better performance, and vice versa for $\uparrow$. w/o attention denotes a variant of raw DiffCharge which discards the multi-head self-attention layer. $T=30$ to $T=70$ change the total number of diffusion steps $T$ from original $T=50$ to other four different configurations.}
\end{tablenotes}
\end{center}
\end{table*}

\subsection{Battery-level Charging Scenario Generation}
1) \textcolor{black}{\textit{Overall performance}: We randomly generate 1,500 charging curves to validate the efficacy of DiffCharge on battery-level scenarios. In Table \ref{overall metrics}, we compare the holistic generation quality of DiffCharge with other three models on three quantitative scores. It is evident that DiffCharge achieves the best grades, which indicates that in contrast to other generative methods, denoising diffusion-based model can produce more diverse fine-grained charging curves with more plausible battery physical dynamics. Besides, we can observe from Fig. \ref{t-sne} that charging curves generated by DiffCharge can better overlap real data in the 2D t-SNE visualization space, which manifests better distributional resemblance and temporal pattern coverage. Based on such overall comparison, we can find that data-driven models are more adept at modeling complex dynamics in EV charging time-series than the prior statistical model. In addition, since the learning paradigm of both VAEs and GANs is to directly transform the Gaussian distribution to original data at once time, whereas DDPMs executes this daunting transition in a gradual step-wise manner, so that it could be more efficient for DiffCharge to capture the complex distribution of charging time-series.} In Fig. \ref{acf}, we randomly showcase a handful of our generated curves. Exhibitions in the upper two rows show DiffCharge can produce realistic charging curves with various temporal modes and different charging duration. In the third row, we can see that the real and generated auto-correlation coefficients (with a range of time lags from 1 to 48 minutes) \cite{DONG2022118387} keep close to each other, indicating that DiffCharge can indeed capture inherent time dependencies in original charging curves. Next, we will take a closer look at unique features in battery charging time-series.

\begin{figure}[!t]
\setlength{\abovecaptionskip}{0.cm}
\centering
\includegraphics[width=0.48\textwidth]{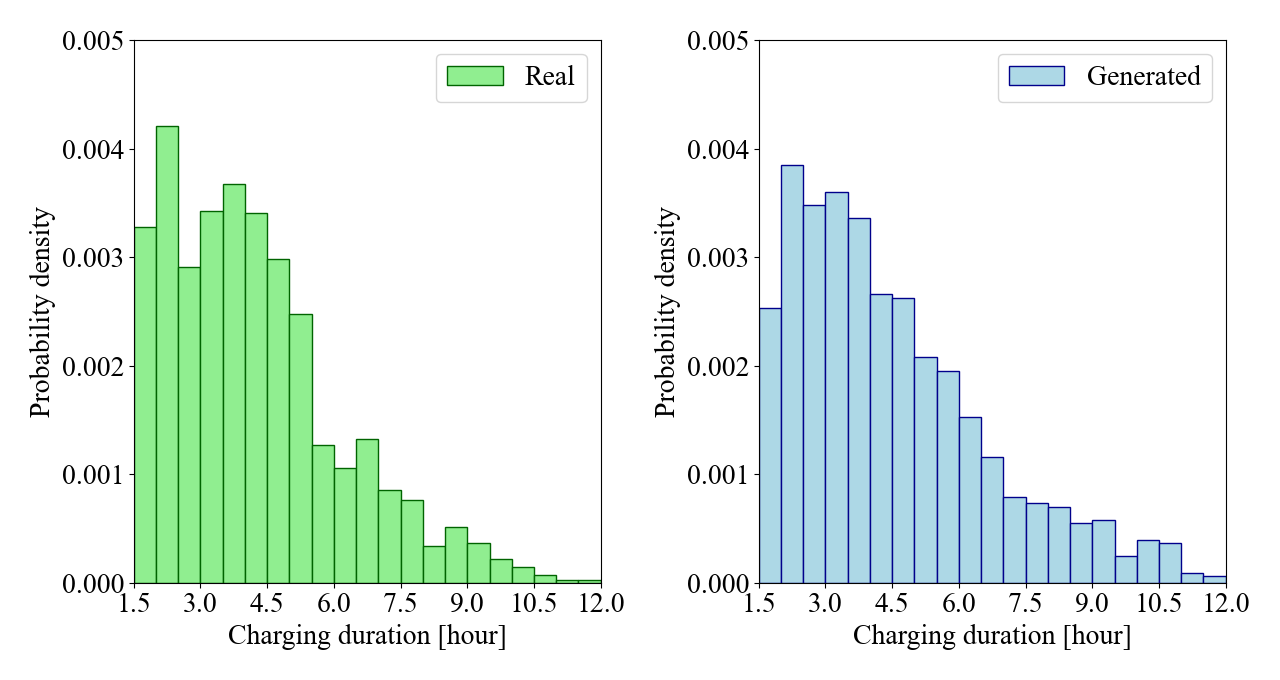}
\caption{Real and generated marginal distribution of charging duration.}
\label{charging duration pdf}
\end{figure}

\begin{figure}[!t]
\centering
\setlength{\abovecaptionskip}{0.cm}
\includegraphics[width=0.48\textwidth]{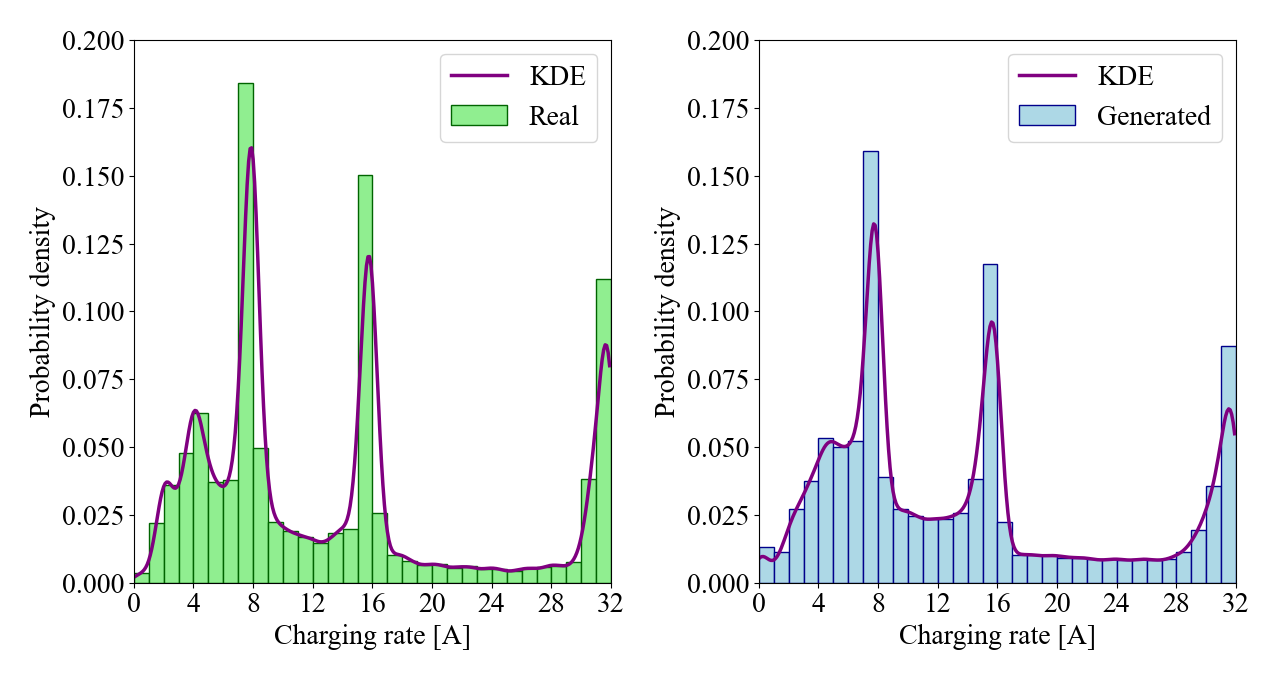}
\caption{Real and generated marginal distribution of charging rates in the bulk stage of battery charging curves.}
\label{charging rate pdf}
\end{figure}

2) \emph{Varying Charging Duration}:  
We truncate zero-padded sub-sequences to assess the valid charging duration of generated curves, and compare its empirical PDF with real one in Fig. \ref{charging duration pdf}. 
In both (a) and (b) of Fig. \ref{charging duration pdf}, the shortest and longest charging duration are 1.5 and 12 hours, and most of charging demands can be fulfilled within 8 hours while a few of EVs need to be charged much longer. Consequently, DiffCharge can realize various lengths of charging duration and maintain a similar duration distribution to real charging sessions.

3) \emph{Temporal Characteristics in the Bulk Stage}: We display the marginal distributions over EV charging rates in the bulk stage for real and generated charging curves in Fig. \ref{charging rate pdf}, which are realized by both kernel density estimation (KDE) \cite{DONG2022118387} and frequency histogram. In both real and generated PDF, there are three notable peak regions located around 8A, 16A and 32A, revealing that in the bulk stage, charging rates primarily follow major trends around these three levels with stochastic fluctuations. As there only exist mild divergences between the real and generated PDF, DiffCharge is able to represent plausible temporal properties in the bulk charging stage.

\begin{figure*}[t]
\centering
\setlength{\abovecaptionskip}{0.cm}
\includegraphics[width=\textwidth]{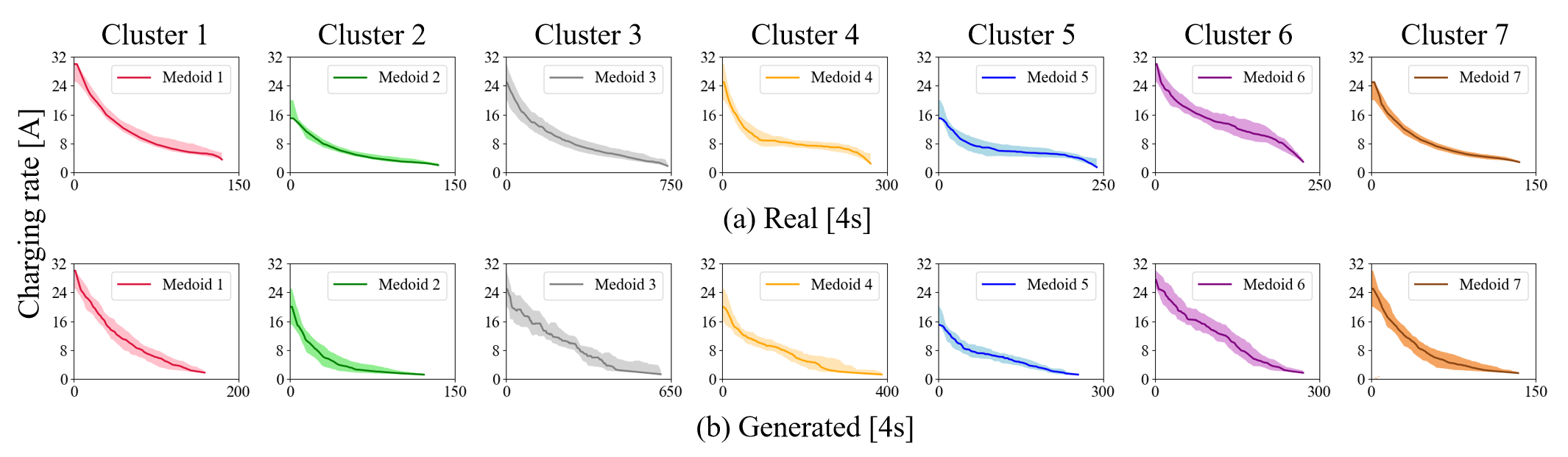}
\caption{\textcolor{black}{Tail feature clustering of the absorption stage for real (upper) and generated (lower) single battery charging curves. Shaded regions represent all analogous tail features within each cluster. Such absorption states are unique in terms of duration and trend.}}
\label{tail clusters}
\end{figure*}

4) \emph{Tail Features in the Absorption Stage}: We utilize the clustering method proposed in \cite{Chenxi2020} to further analyze the specific tail features for generated charging curves. The motivation is that batteries with similar classes of electrochemical properties would exhibit similar tail features, and could be categorized into the same cluster. Hence, we resort to the clustering technique to inspect different types of battery charging dynamics in the absorption stage. \textcolor{black}{We first divide such tail features using K-Means clustering method, and set the empirical cluster number as 7, which can best distinguish each cluster's charging characteristics.} Fig. \ref{tail clusters} displays 7 clusters, and each cluster covers those similar tail features distinguished by absorption stage length, decline speed and etc. Note that the medoid in each cluster signifies the most representative tail feature. 
We can observe that DiffCharge can derive battery electrochemical dynamics by learning on real absorption stages and preserve them in the synthetic charging curves.

\textcolor{black}{\subsection{Supplementary Analysis}
According to the detailed description in Section \ref{sec:model}, we can discover that the multi-head self-attention mechanism and total number of diffusion steps $T$ are two keys to affect both generation quality and computational efficiency of DiffCharge. In this regard, we conduct supplemental analysis for demonstrate the efficacy of such two factors on battery-level scenarios.}

1) \textcolor{black}{\textit{Effectiveness of self-attention}: As claimed in Section \ref{sec:model}-A, as there exist inherent temporal correlations in charging load time-series, we complement the multi-head self-attention to empower the denoising network to learn the complex temporal dynamics and perceive the global temporal patterns. In the fourth row of Table \ref{overall metrics}, we ablate the attention layer in $\bm{\epsilon}_{\theta}$ and it turns out that all of three quantitative scores have dropped conspicuously, which reveals that the utilization of attention mechanism can boost to derive the real temporal properties when modeling the charging time-series distribution. Besides, to investigate additional computation burden after inserting such attention module, we compare with original DiffCharge on the time consumption of generating 1500 samples. As a result, the inference speed (time usage to yield one sample) just increases from 0.11 to 0.17 seconds. Overall, such self-attention component is of vital significance to generate higher quality charging temporal scenarios even though it will sacrifice a bit of computation efficiency.}

2) \textcolor{black}{\textit{Effect of step number $T$}: As stated in \cite{Nichol2021}, the step number $T$ is a key parameter for diffusion models. A larger $T$ can guarantee the reverse transition to approximate Gaussian form, but compound the generating speed as well. In Table \ref{overall metrics}, we present the performance of DiffCharge trained on different values of $T$. We can observe that the inference speed gradually decreases while $T$ is growing, and DiffCharge with $T=50$ can achieve the best outcomes on the specific battery charging curve generation task. When $T$ is relatively small (e.g. 30, 40), prescribing the reverse transition defined in \eqref{eq:6} as Gaussian distribution could be skewed (non-trivial discretized errors of stochastic differential equations, refer to \cite{song2020score} for more explanation) and cause inaccurate modeling for $q(\mathbf{x}_{0})$. While $T$ becomes too large (e.g. 60, 70), some redundant parts by the end of noise scheduling will happen and do not contribute the generation quality \cite{Nichol2021}. How to quantify the effect of $T$ on learning diffusion models and efficiently determine the proper $T$ remains a critical issue, we leave it for future research.}

\begin{table}[!t]
\begin{center}
\caption{\textcolor{black}{Quantitative evaluation on conditional generation.}\vspace{-8pt}}
\setlength{\abovecaptionskip}{0.cm}
\label{station metrics}
\begin{tabular}{ccc}
\hline\\[-2.9mm]\hline
Station & Marginal Score & Discriminative Score \\ \hline
JPL     & 0.0169         & 0.6712±0.0349        \\
Caltech & 0.0367         & 0.6452±0.0061        \\ 
\hline\\[-2.9mm]\hline
\end{tabular}
\end{center}
\end{table}

\begin{figure}[!t]
\centering
\setlength{\abovecaptionskip}{0.cm}
\includegraphics[width=0.45\textwidth]{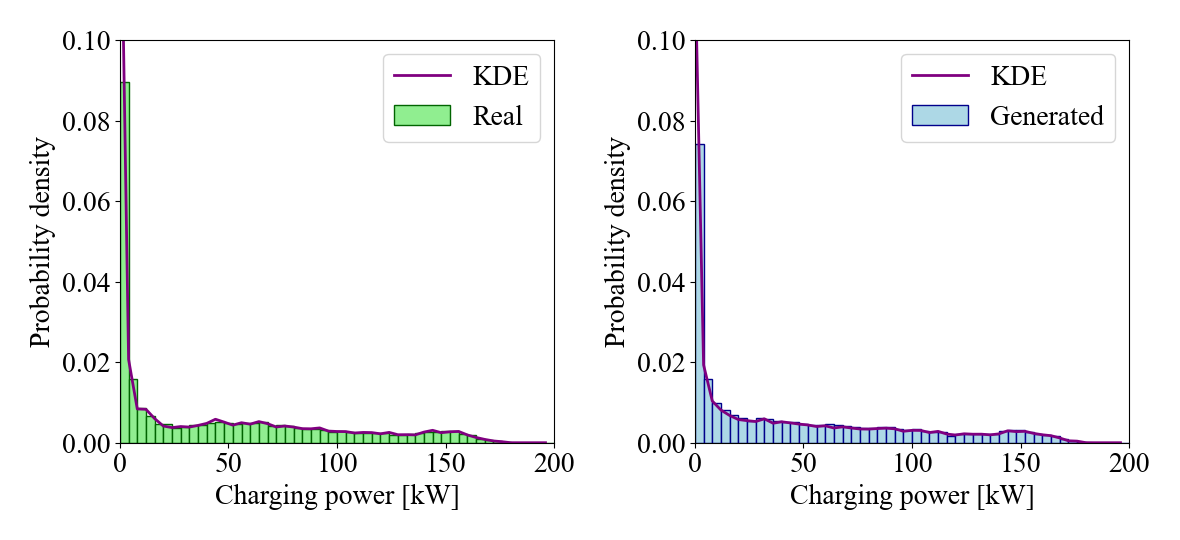}
\caption{Real and generated marginal distribution of JPL charging load.}
\label{jpl pdf}
\end{figure}

\begin{figure}[!t]
\centering
\setlength{\abovecaptionskip}{0.cm}
\includegraphics[width=0.45\textwidth]{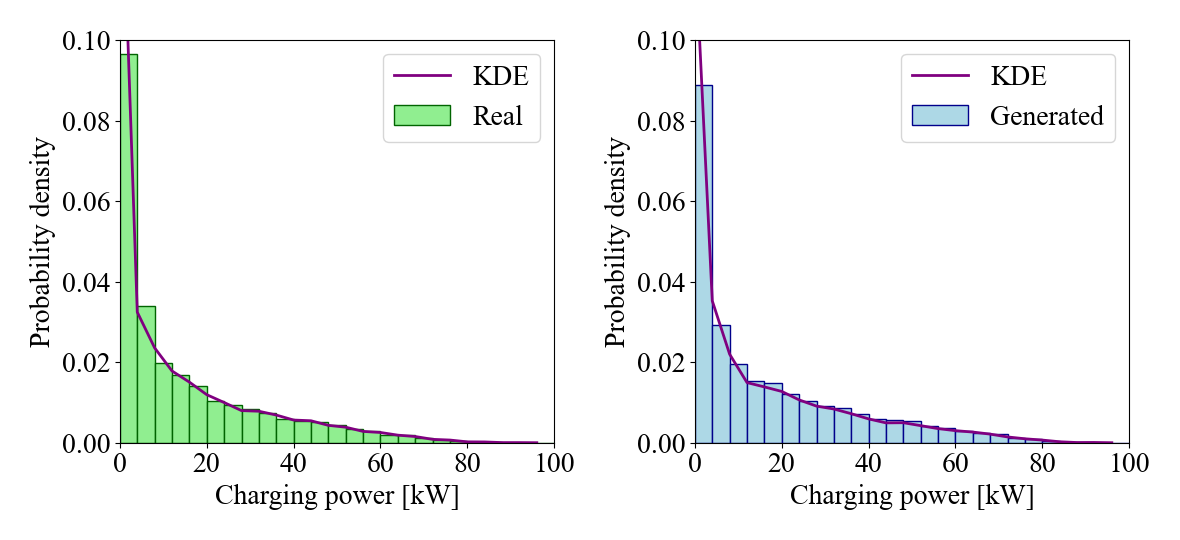}
\caption{Real and generated marginal distribution of Caltech charging load.}
\label{caltech pdf}
\end{figure}

\subsection{Station-level Charging Scenario Generation}
As for station charging demand, we focus on the station-conditional generation, namely producing distinct patterns of charging load profiles for different stations. In our execution, we feed the historical daily charging load time-series and their corresponding labels of both JPL and Caltech stations to DiffCharge during the training stage. Note that the label $y_{s}$ is a two-dimensional one-hot vector, which signifies a specific class of charging station (JPL is a workplace station whereas Caltech is a campus station). In evaluation, we employ the trained DiffCharge to generate a batch of charging load profiles for both stations, and validate whether these two batches of generated samples can realize the realistic and unique temporal patterns of their respective station types. \textcolor{black}{In Table \ref{station metrics}, we exhibit quantitative evaluation results of generated profiles for both charging stations. It is obviously that two marginal scores are small enough and discriminative scores are close to 0.7, which implies the great conditional distribution similarity. In Fig. \ref{jpl pdf} and Fig. \ref{caltech pdf}, we also show the marginal distribution of real and generated charging power for two stations. Evidently, the generated PDF of charging load notably resembles the real PDF, and the charging load distribution of such two kinds of stations are quite disparate. It attests that DiffCharge is able to procure the accurate conditional distribution $q(\mathbf{s}|y_{s})$.}

\begin{figure}[!t]
\centering
\setlength{\abovecaptionskip}{0.cm}
\includegraphics[width=0.45\textwidth]{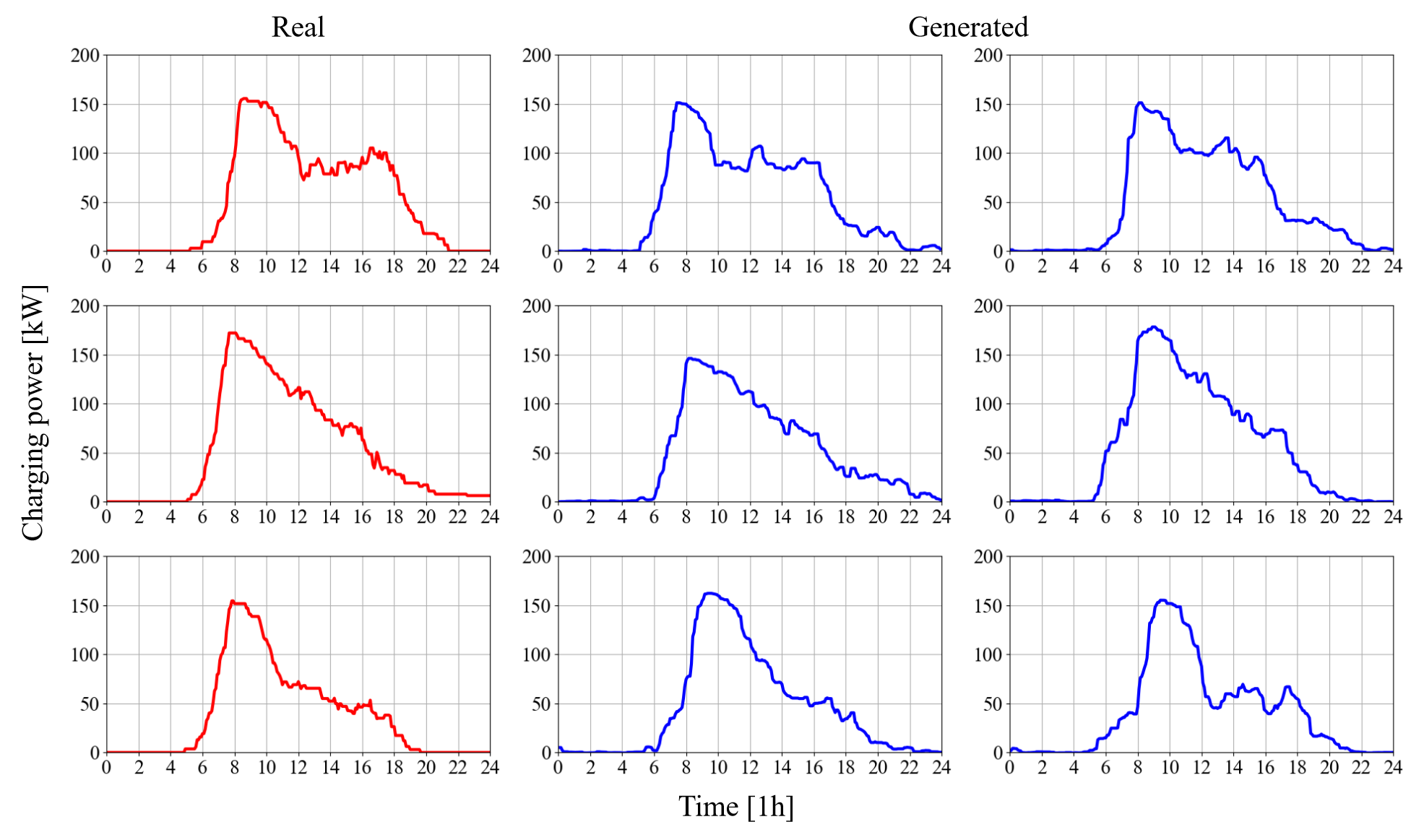}
\caption{Three representative patterns of JPL charging load profiles, while each row indicating a typical mode.}
\label{jpl profiles}
\end{figure}

\begin{figure}[!t]
\centering
\setlength{\abovecaptionskip}{0.cm}
\includegraphics[width=0.45\textwidth]{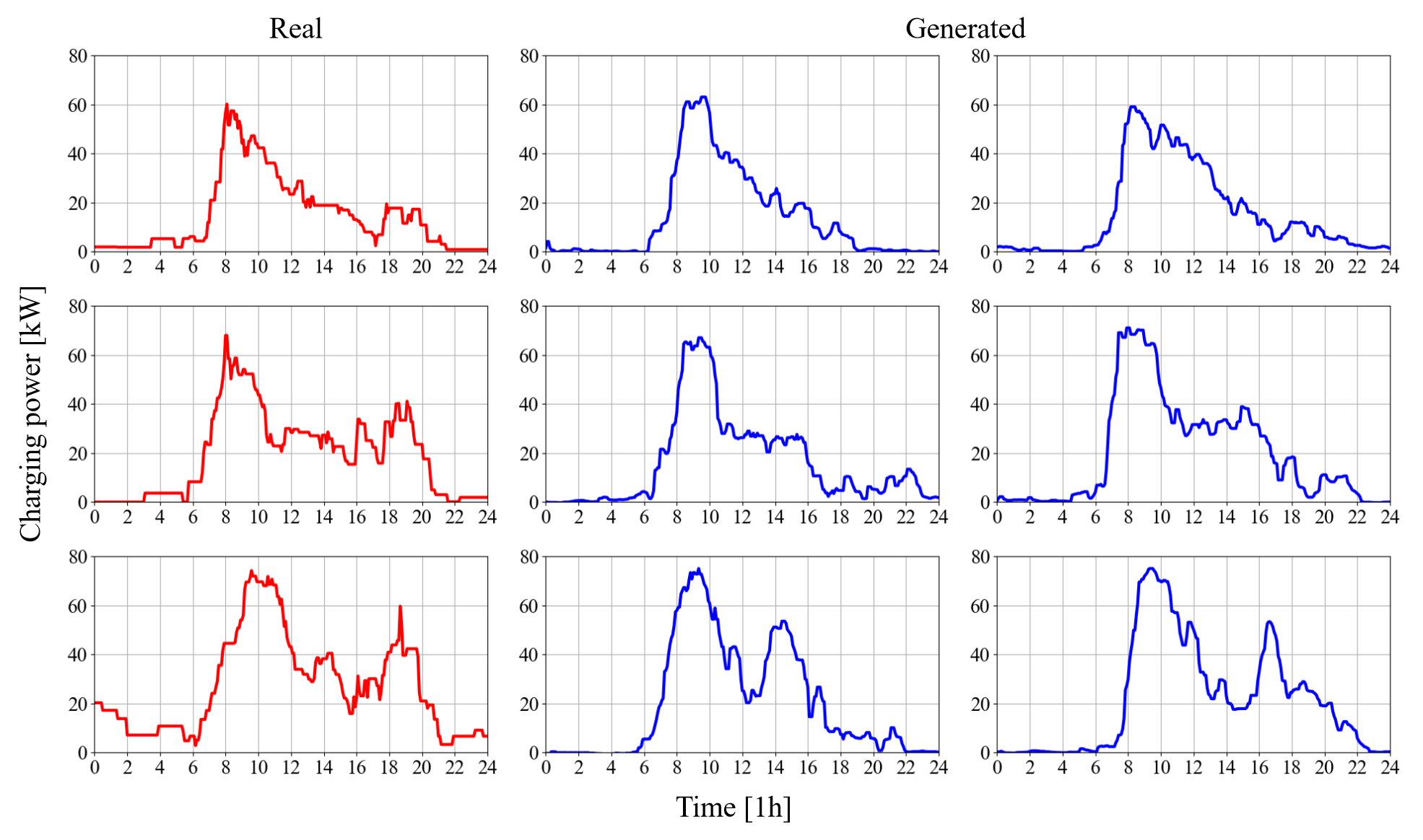}
\caption{Three representative patterns of Caltech charging load profiles, while each row indicating a typical mode.}
\label{caltech profiles}
\end{figure}

Furthermore, we showcase three representative temporal modes of charging load time-series for each station in Fig. \ref{jpl profiles} and Fig. \ref{caltech profiles}. Each red profile indicates a typical class of real charging demand and usage pattern, and the two follow-up blue profiles are generated samples which exhibit similar temporal patterns. The usage pattern differences between two stations arise from different EV charging habits. The biggest difference is that there could be one more peak load in evening at Caltech station, while this situation does not happen at JPL station. Since JPL station is located at a workplace, the majority of charging demands distribute within the working time. Many staff choose to charge their EVs when getting the office and cease charging when departing from the workplace. While Caltech is a campus station, apart from some drivers who work at university, there also exist some users living on campus. These results verify that by utilizing the conditional DiffCharge, it is convenient to yield a broad variety of unique charging load profiles for different stations and apply them to potential station planning.

\begin{figure}[!t]
\centering
\setlength{\abovecaptionskip}{0.cm}
\includegraphics[width=0.45\textwidth]{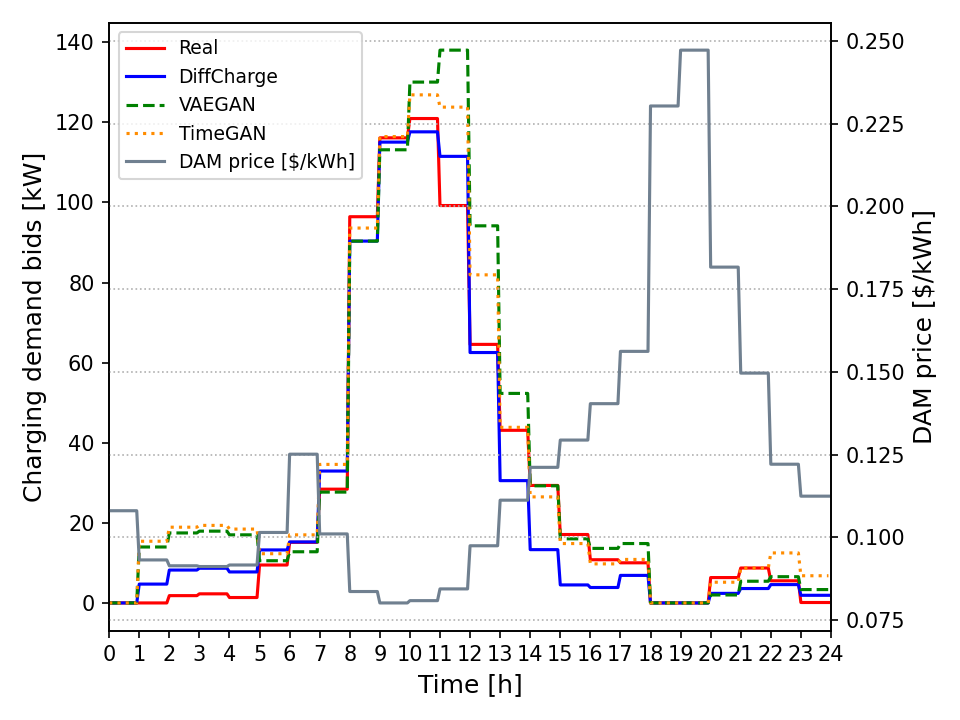}
\caption{\textcolor{black}{Day-head charging demand bids of real session data and simulating scenarios generated by different models on August 30, 2019 at JPL site.}}
\label{demand bids}
\end{figure}

\begin{table}[!t]
\begin{center}
\caption{\textcolor{black}{Bidding costs in different simulated scenarios.}\vspace{-8pt}}
\setlength{\abovecaptionskip}{0.cm}
\label{bid costs}
\begin{tabular}{ccccc}
\hline\\[-2.9mm]\hline
Bidding Costs {[}\${]} & Real    & DiffCharge & VAEGAN  & TimeGAN \\ \hline
Energy Procurement    & 64.8104 & 59.9983    & 77.5166 & 76.9242 \\
User Penalty          & 2.7472  & 4.3787     & 6.9442  & 7.7025  \\
Total                 & 67.5576 & 64.377     & 84.4608 & 84.6267 \\
\hline\\[-2.9mm]\hline
\end{tabular}
\end{center}
\end{table}

\subsection{\textcolor{black}{Charging Energy Bidding in Day-head Market}}
\textcolor{black}{To further demonstrate the usefulness of synthetic charging scenarios for real-world EV integration, we apply them to a practical problem of EV charging energy bidding in day-head market (DAM). Such formulation was initially employed in \cite{jahangir2021novel} to verify the efficacy of data-driven EV demand modeling. In this situation, the station operator's objective is to determine the optimal day-ahead bidding plan of charging power $p_{n,i}$ for each EV $n\in \mathcal{N}$ at time $i\in \mathcal{T}$, which can simultaneously minimize the total charging energy procurement cost $C_{s}$ and maximize charging session satisfaction for every EV user. We use $\mathcal{N}$ and $\mathcal{T}$ to denote the set of EVs and time steps respectively. The latter part is committed to fulfilling individual EV charging demand and can be prescribed as the user penalty cost $C_{u}$. The problem formulation is presented below:}
\textcolor{black}{\begin{subequations}
\begin{align}
\minimize_{p_{n,i}}  \quad & C_{s}+C_{u} \\
s.t. \quad & C_{s}=\sum_{i\in \mathcal{T} }^{} \rho _{i}^{DAM}\sum_{n\in \mathcal{N} }^{} p_{n,i}\Delta i \\
& C_{u}=(G_{f}+G_{d})\rho^{PEN} \\
& G_{f}=\sum_{n\in \mathcal{N} }^{}\sum_{i\in \mathcal{T} }^{}  \left \| p_{n,i}-d_{n,i} \right \| _{2}^{2} \\
& G_{d}=\sum_{n\in \mathcal{N} }^{} \left \| \sum_{i\in \mathcal{T} }^{}(p_{n,i}-d_{n,i})  \right \| _{2}^{2} \\
& 0\le p_{n,i}\le CR_{n}, \; \forall n\in \mathcal{N}, i\in \mathcal{T}
\end{align}
\end{subequations}}

\textcolor{black}{In this formulation, $C_{s}$ in (22a) is defined as the total cost of charging energy bids for all EVs over each time interval $\Delta i$ under DAM electricity price $\rho_{i}^{DAM}$ [\$/kWh]. $C_{u}$ in (22b) denotes the penalty cost on the unexpected gap between planned $p_{n,i}$ and actual demand $d_{n,i}$ with the penalty price $\rho^{PEN}$. Such power gap consists of $G_{f}$ in (22d) and $G_{d}$ in (22e). $G_{f}$ indicates $p_{n,i}$ needs to tightly track the real battery charging curve $d_{n,i}$, whilst $G_{d}$ reflects we also want to fulfill the total charging demand of each EV $n$. Note that \cite{jahangir2021novel} only incorporated the total demand $\sum_{i\in \mathcal{T} }^{}d_{n,i}$ of each charging session $n$, while in our case, we can offer the fine-grained charging demand time-series $d_{n,i}$ with detailed temporal states. We can utilize the generated battery charging curves to simulate the unseen $d_{n,i}$. (22f) is a physical constraint on the maximum charging power of each EV battery.}

\textcolor{black}{We take a workday on Friday, August 30, 2019 at JPL station with 28 EVs as the test case. For real $d_{n,i}$ simulation, we apply our trained DiffCharge to generate an array of battery charging curves and equip each curve with a random arrival time drawn from its estimated distribution. Then, we employ K-Means as the scenario reduction method to obtain the target number of representative scenarios. The DAM electricity price $\rho_{i}^{DAM}$ can be fetched from CAISO website and the user penalty price $\rho^{PEN}$ is set to 0.8 times of maximum $\rho_{i}^{DAM}$. The time interval $\Delta i$ is 5 minutes, and the charging power capacity $CR_{n}$ is 10kW in ACN-data. We employ CVXPY toolkit to directly solve this convex problem. To validate the quality of simulated synthetic scenarios, we yake real charging sessions on the test day as ground-truth signal, and the optimized $p_{n,i}$ incurred by realistic synthetic $d_{n,i}$ is supposed to approach the bidding scheme in the real situation.}

\textcolor{black}{In Fig. \ref{demand bids}, we depict the real-time charging demand bids (total power bids over all EVs at each time $i$) of the true scene and simulated scenarios produced by three generative models. Apparently, the bidding plan elicited by DiffCharge can best follow the actual situation. Besides, we list two kinds of bidding cost $C_{s}$ and $C_{u}$ in Table \ref{bid costs}, and consequently, costs calculated on diffusion-based scenarios is closest to real costs. Such application results further demonstrate that the usefulness of our generated EV charging scenarios.}

\section{Conclusion}
\label{sec:conclusion}
In this paper, we design a novel denoising diffusion model termed DiffCharge to generate diverse and realistic EV charging scenarios. The proposed model is capable of accurately approximating the complex distribution of real EV charging time-series data, addressing the charging uncertainty quantification issue for EV planning and operation. The multi-head self-attention mechanism is exploited to learn the underlying temporal correlations for both battery-level and station-level time-series scenarios. DiffCharge can not only derive volatile temporal dynamics in various battery charging curves, but also yield distinct modes of charging load profiles for different charging stations. Evaluation results demonstrate our generative method is able to efficiently derive charging uncertainties associated with large-scale EV integration. In future research, we will further explore the controllable generation capability of diffusion models to produce specific charging scenarios towards variable conditions such as initial SOC, battery types and station congestion situations. Miscellaneous EV-enabled contexts like demand response, vehicle-to-grid optimization will be also considered for generating future charging data.


\bibliographystyle{IEEEtran}
\bibliography{reference}

\end{document}